\documentclass[fleqn,usenatbib]{mnras}

\usepackage{newtxtext,newtxmath}

\usepackage[T1]{fontenc}

\usepackage{booktabs}

\DeclareRobustCommand{\VAN}[3]{#2}
\let\VANthebibliography\thebibliography
\def\thebibliography{\DeclareRobustCommand{\VAN}[3]{##3}\VANthebibliography}

\usepackage{graphicx}	
\usepackage{amsmath,siunitx}	
\usepackage{subfig}
\usepackage[normalem]{ulem}

\title[Long-term evolution of PSR~J0901$-$4046]{Slow and steady: long-term evolution of the 76-second pulsar~J0901$-$4046}

\author[M. C. Bezuidenhout et al.]{M.~C.~Bezuidenhout,$^{1}$\thanks{E-mail: tiaan.bezuidenhout@skao.int}
N.~D.~R.~Bhat$^{2}$,
M.~Caleb$^{3,4}$,
L.~N.~Driessen$^{5}$,
F.~Jankowski$^{6}$,
M.~Kramer$^{7,8}$,\newauthor
V.~Morello$^{9}$,
I. Pastor-Marazuela$^{8}$,
K.~Rajwade$^{10}$,
J. Roy$^{11}$, 
B.~W.~Stappers$^{8}$,
M.~Surnis$^{12}$,
and J. Tian$^{8}$
\\
$^{1}$Department of Mathematical Sciences, University of South Africa, Cnr Christiaan de Wet Rd and Pioneer Avenue, Florida Park, 1709, Roodepoort, South Africa\\
$^{2}$International Centre for Radio Astronomy Research, Curtin University, Bentley, WA 6102, Australia\\
$^{3}$ARC Centre of Excellence for Gravitational Wave Discovery (OzGrav), Hawthorn, 3122, Victoria, Australia\\
$^{4}$CSIRO, Space and Astronomy, PO Box 1130, Bentley, WA 6102, Australia\\
$^{5}$Sydney Institute for Astronomy, School of Physics, The University of Sydney, NSW 2006, Australia\\
$^{6}$LPC2E, OSUC, Univ Orleans, CNRS, CNES, Observatoire de Paris, F-45071 Orleans, France\\
$^{7}$Max-Planck-Institut für Radioastronomie, Bonn, Germany\\
$^{8}$Jodrell Bank Centre for Astrophysics, Department of Physics and Astronomy, The University of Manchester, Oxford road, Manchester, M13 9PL, United Kingdom\\
$^{9}$SKA Observatory, Jodrell Bank, Lower Withington, Macclesfield, Cheshire, SK11 9FT, UK\\
$^{10}$Astrophysics, University of Oxford, Denys Wilkinson Building, Keble road, Oxford OX1 3RH, UK\\
$^{11}$National Centre for Radio Astrophysics, Tata Institute of Fundamental Research, Pune 411007, India\\ 
$^{12}$Department of Physics, Indian Institute of Science Education and Research Bhopal, Bhopal Bypass Road, Bhauri Bhopal 462 066 Madhya Pradesh, India 
}

\date{Accepted XXX. Received YYY; in original form ZZZ}

\pubyear{2024}

\begin{document}
\defcitealias{2022caleb}{C22}

\label{firstpage}
\pagerange{\pageref{firstpage}--\pageref{lastpage}}
\maketitle

\begin{abstract}
PSR~J0901$-$4046, a likely radio-loud neutron star with a period of 75.88~seconds, challenges conventional models of neutron star radio emission. Here, we showcase results from 46 hours of follow-up observations of PSR~J0901$-$4046 using the MeerKAT, Murriyang, GMRT, and MWA radio telescopes. We demonstrate the intriguing stability of the source's timing solution over more than three years, leading to an RMS arrival-time uncertainty of just $\sim$10$^{-4}$ of the rotation period. Furthermore, non-detection below 500~MHz may indicate a low-frequency turnover in the source's spectrum, while no secular decline in the flux density of the source over time, as was apparent from previous observations, has been observed. Using high time-resolution MeerKAT data, we demonstrate two distinct quasi-periodic oscillation modes present in single pulses, with characteristic time scales of 73~ms and 21~ms. We also observe a statistically significant change in the relative prevalence of distinct pulse morphologies compared to previous observations, possibly indicating a shift in the magnetospheric composition over time. Finally, we show that the W$_{50}$ pulse width is nearly constant from 544--4032~MHz, consistent with zero radius-to-frequency mapping. The very short duty cycle ($\sim$1.4$^{\circ}$) is more similar to radio pulsars with periods $>$5~seconds than to radio-loud magnetars. This, along with the lack of magnetar-like outbursts or timing glitches, complicates the identification of the source with ultra-long period magnetar models.
\end{abstract}




\section{Introduction}

Pulsars are rapidly rotating neutron stars (NSs) that are observed as emitters of pulsed radio waves when coherent radiation beamed from their magnetic poles sweeps across an observer's line of sight. They rotate with periods ($P$) ranging from milliseconds to tens of seconds. Recently, an ultra-long period ($\sim$75.88~s) pulsar was discovered \citep[][hereafter C22]{2022caleb} in a real-time single pulse search carried out by the Meer (More) TRAnsients and Pulsars \citep[MeerTRAP,][]{2018sanidas, 2020jankowski, 2020malenta, 2021rajwade}\footnote{\url{https://www.meertrap.org}} and ThunderKAT\footnote{\url{http://www.thunderkat.uct.ac.za}} \citep[][]{2016fender} projects using the MeerKAT radio telescope in South Africa \citep{2016jonas}.

The source, PSR~J0901$-$4046, has a remarkable combination of characteristics that sets it apart from all other pulsars discovered so far. Chiefly, its period ($P = 75.88$~s) and period derivative ($\Dot{P} = 2.25 \times 10^{-13}$) imply a surface magnetic field strength $B = 3.2\times10^{19}\sqrt{P\Dot{P}} \simeq 1.3\times10^{14}$~G, which exceeds the quantum critical field $B_{\mathrm{cr}} = m_{e}^{2}c^{3}/ \left(e \hbar \right)=  4.413\times10^{13}$~G. Above this limit, the electron cyclotron energy exceeds its rest mass energy, and quantum-electrodynamical processes become important \citep{2005heyl}. It has been theorised that the single-photon pair production ($\gamma\rightarrow e^{+} + e^{-}$) that gives rise to radio emission in pulsar magnetospheres is suppressed in the presence of ultra strong fields exceeding $B_{\mathrm{cr}}$ \citep[][]{1998baring}.

Its strong magnetic field may suggest that PSR~J0901$-$4046 is a magnetar, a subclass of pulsars with extremely strong magnetic fields. Magnetars are characterised by the dramatic variability of their emission, featuring short, millisecond-duration bursts to major month-long outbursts. However, magnetars are commonly young \citep[most $\lesssim10^{6}$ years;][]{2014olausen}{}{}\footnote{\url{http://www.physics.mcgill.ca/~pulsar/magnetar/main.html}}, while PSR~J0901$-$4046 has a relatively large characteristic age,  $\tau = P/2\Dot{P} \simeq 5.3\times10^{6}$~years. Magnetars are also most often only visible at X-ray and $\gamma$-ray wavelengths, while follow-up observations of PSR~J0901$-$4046 with the \textit{Swift}/XRT telescope have not produced any detections in the 2--10~keV range.  The inferred upper limit on the X-ray luminosity at 0.5--2~keV of L$_{X} \lesssim 3.2 \times 10^{30}$ erg~s$^{-1}$ exceeds the pulsar's rotational energy loss $\dot{E} \simeq 2.0 \times 10^{28}$ erg~s$^{-1}$, indicating that it may still be a magnetar \citepalias{2022caleb}.

Instances have been recorded of magnetars exhibiting radio emission following a high-energy outburst, which dissipates over time \citep[e.g.][]{2006camilo,2007camilo,2020esposito}. Only one magnetar has been discovered via its radio emission, with no apparent high-energy outburst observed \citep{2010levin}. Given its advanced age and the lack of an associated high-energy outburst, it is uncertain how PSR~J0901$-$4046 fits into the existing picture of magnetar radio emission. \citetalias{2022caleb} found that the mean PSR~J0901$-$4046 pulse brightness is possibly declining secularly on short time scales, which would be consistent with a radio-loud magnetar transitioning into radio quiescence. Previous long-term radio-wavelength observations of radio magnetars have revealed erratic variations in pulsed flux density over time, including periods of quiescence, typically accompanied by timing glitches or the appearance of giant pulses \citep[][]{2018camilo,2022rajwade,2022caleb}. However, this radio variation has appeared uncorrelated with the observed secular decline in these magnetars' X-ray flux, which has been interpreted as thermal emission from a hot spot on the surface of the NS that is shrinking \citep[see][]{2020hu,2022rajwade}.

PSR~J0901$-$4046 also displays unusual temporal behaviour. Although magnetar pulses are noted for their sub-pulse structure \citep[see][]{2017kaspi}, the structural features of PSR~J0901$-$4046's single pulses vary drastically from pulse to pulse. In \citetalias{2022caleb}, the profiles are divided into seven distinct types. PSR~J0901$-$4046's emission seems to vacillate at random between these types. Additionally, some bright pulses' sub-pulse components seem to repeat quasi-periodically (every 9.57 to 338 ms), behaviour common to all known radio-loud magnetars \citep[][]{2024kramer}{}{}, as well a few Fast Radio Bursts (FRBs) to date \citep[][]{2021majid,2021chime,2023Pastor}{}{}. Furthermore, the most common quasi-periodicity is $\sim76~\mathrm{ms} \sim (P/s)\times10^{-3}$, a relationship similar to what has been observed in quasi-periodic ``micropulses'' in some radio pulsars \citep[][]{1979cordes} and millisecond pulsars \citep[][]{2016de}{}{}. \citet{2024kramer} suggests that quasi-periodic substructure with a $P\times10^{-3}$ scaling is a common feature of all radio-emitting NSs, regardless of evolutionary history. This points to a universal mechanism responsible for the substructure involving charge packets accelerated to the same $\gamma$-factor, such that their emitting time scales are constrained by relativistic beaming dependent only on the star's angular velocity \citep[][]{2024kramer}.

Recently, a class of mature, highly magnetised NSs called Ultra-Long Period Magnetars (ULPMs) has been theorised as a progenitor for extragalactic FRBs \citep[][]{2020beniamini}. Radio transients exhibiting prolonged periods ranging from 7 minutes to 6.25 hours have been discovered in recent years that may belong to this class \citep[][]{2022hurleywalker,2023hurleywalker,2024hurleywalker,2024dong,2024caleb,2024deruiter,2025lee}. PSR~J0901$-$4046's parameters are also consistent with those postulated for ULPMs. It may be that these kinds of transients provide a direct connection between magnetars and FRBs, which has often been hypothesised to exist \citep[e.g.][]{2017metzger,2020chime,2020bochenek,2019cordes}. Note that instances of pulsar-like emission with minute--hour periodicity have been observed from a few binary white dwarf systems \citep[][]{2016marsh,2023pelisoli,2024deruiter,2025rodriguez}, but without detection of a companion or any evidence for binarity from the timing (see $\S\ref{subsec:timing}$), we assume a NS origin in this study.

In this paper we present results from long-term follow-up observations of PSR~J0901$-$4046 with the MeerKAT, Murriyang, GMRT, and MWA radio telescopes (see $\S\ref{sec:obs}$), including analysis of its timing solution, pulse energy, and pulse morphology ($\S\ref{sec:results}$). 

\section{Observations and Analysis}\label{sec:obs}

\begin{figure*}
    \centering
    \includegraphics[width=\textwidth]{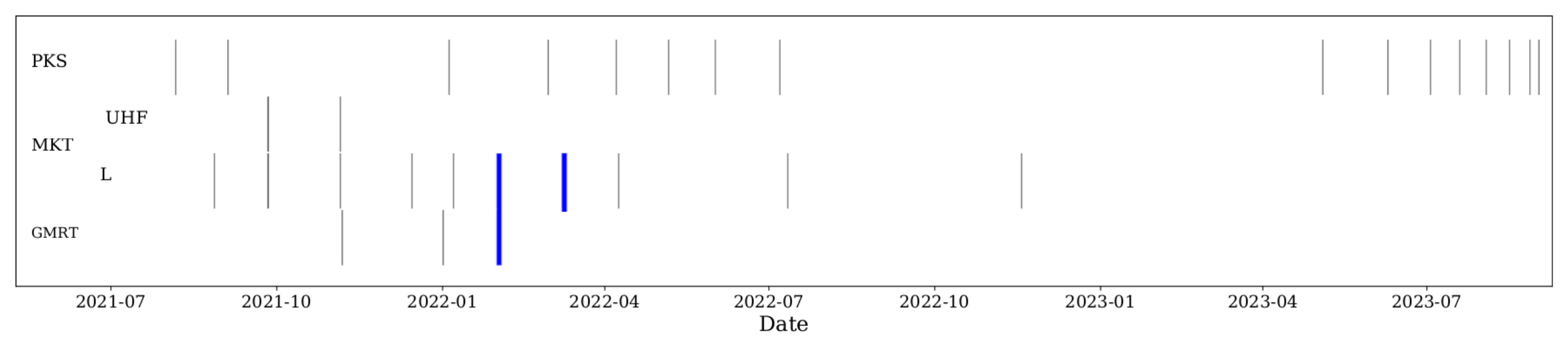}
    \caption[Follow-up observing dates of PSR~J0901$-$4046.]{Observation dates of PSR~J0901$-$4046 using the GMRT radio telescope, the MeerKAT radio telescope's UHF-band and L-band receivers, and the Murriyang radio telescope's UWL receiver. The blue lines indicate the dates of simultaneous multi-instrument observations. The second simultaneous observation included an MWA observation at 140--170~MHz that did not produce any detections of the pulsar.}
    \label{fig:obsdates}
\end{figure*}

We have performed a total of 46.25 hours of follow-up observations of PSR~J0901$-$4046 using four different radio telescopes, viz.~MeerKAT, the 64~metre Murriyang telescope at Parkes Observatory in Australia, the Giant Metrewave Radio Telescope (GMRT) in India, and the Murchison Widefield Array (MWA) in Australia, from August 2021 to May 2024. This includes a coordinated simultaneous hour-long observation of PSR~J0901$-$4046 in both the MeerKAT L-band (using the FBFUSE and TUSE clusters) and GMRT Band-3 and Band-4 on 31 January 2022, as well as a simultaneous two-hour observation on 12 March 2022 using MeerKAT and the MWA. Fig.~\ref{fig:obsdates} shows the dates of these observations, with the date of the multi-telescope observations indicated in blue. The observational parameters are listed in Table~\ref{table:obsparam}.

\subsection{MeerKAT data}\label{subsec:meerkatobs}
MeerKAT observations of PSR~J0901$-$4046 were carried out across twelve occasions amounting to 20 hours on source. On two occasions, the observations were performed in the sub-array mode for simultaneous L-band (856--1712~MHz) and UHF-band (544--1088~MHz) observations, while all other observations were carried out using the L-band receiver only.

\begin{table*}
\centering
\caption[PSR~J0901$-$4046 follow-up observation parameters.]{Observational parameters for all observations of PSR~J0901$-$4046.}\label{table:obsparam}
\begin{tabular}{llccclcll}
\toprule
Date & Start UTC & Length (min) & Rotations & Telescope(s) & Band(s) & MKT antennas & N$_{\mathrm{chan}}$\\
\midrule
2021/08/05 & 21:14:09 & 60 & 47 & PKS & UWL & & 3328\\
2021/08/27 & 10:08:59 & 30 & 23 & MKT & L + UHF & 56 & 4096 + 4096\\
2021/09/03 & 21:06:68 & 120 & 94 & PKS & UWL & & 3328\\
2021/09/26 & 01:06:09 & 420 & 331 & MKT & L + UHF & 56 &  4096 + 4096\\
2021/11/05 & 06:00:00 & 60 & 47 & MKT & L  & 56 & 1024\\
2021/11/06 & 01:50:22 & 120 & 94 & GMRT & Band-3 + Band-4 & & 4096 + 4096\\
2021/12/12 & 23:06:40 & 120 & 47 & MKT & L  & 52 &  1024 \\
2021/12/14 & 23:53:16 & 60 & 94 & GMRT & Band-3 + Band-4  & 52 &  4096 \\
2022/01/01 & 18:54:46 & 180 & 142 & GMRT & Band-3 + Band-4 & & 4096 + 4096\\
2022/01/04 & 13:49:03 & 60 & 47 & PKS & UWL & & 3328\\
2022/01/07 & 00:44:36 & 60 & 47 & MKT & L  & 60 & 1024\\
2022/01/31 & 19:44:41 & 60 & 47 & MKT + GMRT & L +  Band-3 + Band-4 & 56  & 1024 + 4096 + 4096\\
2022/02/28 & 13:08:40 & 105 & 83 & PKS & UWL & & 3328\\
2022/03/12 & 15:30:00 & 120 & 94 & MKT + MWA & L & 56 & 1024 + 1024\\
2022/04/07 & 08:25:13 & 30 & 24 & PKS & UWL & & 3328 \\
2022/04/08 & 16:18:27 & 60 & 47 & MKT & L & 64 & 1024\\
2022/05/06 & 09:07:07 & 45 & 35 & PKS & UWL & & 3328\\
2022/06/01 & 06:40:05  &  15 & 12 & PKS & UWL & & 3328\\
2022/07/07 &  01:38:37 &  15 & 12 & PKS & UWL & & 3328\\
2022/07/21 &   11:53:18  &45& 38 & MKT & L & 56  & 4096\\
2022/11/18 &   03:30:52 &100 & 78 & MKT & L & 64 & 1024\\
2023/01/19 & 18:00:37 & 60 & 51 & MKT & L & 60 & 1024\\
2023/05/04 &  05:59:50 & 75 & 59 & PKS & UWL & & 3328\\
2023/05/19 & 07:03:38 & 30 & 24 & PKS & UWL & & 3328 \\
2023/06/09 &   08:35:53 & 60& 49 & MKT & L & 56 & 1024\\
2023/06/09 &   09:00:38 & 110& 87 & PKS & UWL & & 3328\\
2023/06/11 &   02:10:35 & 60 & 49 & PKS & UWL & & 3328\\
2023/07/02 &   23:42:54 & 60  & 49 & PKS & UWL & & 3328\\
2023/07/19 &  06:09:29 & 45  & 35 & PKS & UWL & & 3328\\
2023/07/31 &  21:41:37 & 60  & 51 & PKS & UWL & & 3328\\
2023/08/11 &  05:49:58 & 60  & 51 & PKS & UWL & & 3328\\
2023/08/15 &  21:09:59 & 15  & 12 & PKS & UWL & & 3328\\
2023/08/27 & 03:43:24 &  45 & 34 & PKS & UWL & & 3328\\
2023/09/05 & 03:19:13 & 90  & 68 & PKS & UWL & & 3328\\
2023/09/06 & 23:43:50 & 120  & 94 & PKS & UWL & & 3328\\

\midrule
\textbf{Total} & & 46.25 hrs. & 2196\\
\bottomrule
\end{tabular}
\end{table*}

The dual sub-array observations on 27 August and 26 September 2021 were performed using the Pulsar Timing User Supplied Equipment (PTUSE) backend of the MeerKAT pulsar timing project MeerTime described in \citet[][]{2020bailes}. These data were recorded in the PTUSE search mode with a sampling time of $\sim$301~$\upmu$s in the \textsc{psrfits} format \citep[][]{2004Hotan}. Both the UHF-band and L-band data are split into 4096 frequency channels and coherently de-dispersed at a DM of 52.6~pc~cm$^{-3}$. Four polarisation channels were recorded.

On the subsequent dates, the MeerKAT observations were performed using the Filterbank BeamFormer User Supplied Equipment (FBFUSE) cluster \citep[][]{2021chen} with a sampling time of $\sim$306~$\upmu$s. On each occasion, FBFUSE used the raw voltage data from 52 to 64 MeerKAT dishes to form beams, and filterbank data were recorded to the Accelerated Pulsar Search User-Supplied Equipment (APSUSE) cluster \citep[][]{2023Padmanabh}.  These data were re-channelised into 1024 frequency channels and stored in the \texttt{SIGPROC}\footnote{\url{http://sigproc.sourceforge.net/}} \texttt{filterbank} format \citep[][]{2011lorimer}. The FBFUSE data were not coherently de-dispersed.

The PTUSE and FBFUSE data were folded using the timing solution from \citetalias{2022caleb} and the \texttt{DSPSR} processing package \citep[][]{2011vanstraten} to produce single-pulse stacks in the \textsc{PSRFITS} format. For each hour of observations, the corresponding single pulse stack was cleaned of Radio Frequency Interference (RFI) using \texttt{clfd} \citep{2019morello} as well as the \textsc{psrchive} \citep[][]{2004Hotan} tool \texttt{pazi} to produce an hourly frequency channel mask that could be applied to individual pulse archives. The resulting cleaned single pulse archives were then manually inspected to identify positive detections of PSR~J0901$-$4046. The FBFUSE single pulse data were used for pulsar timing (see $\S$\ref{subsec:timing}) and pulse width analyses ($\S$\ref{subsec:pwidth}), while the PTUSE data were used for the pulse energy ($\S\ref{sec:penergy}$) and morphology analyses ($\S\ref{subsec:pmorph}$).

\subsection{Murriyang UWL data}\label{subsec:pksobs}
We performed observations of PSR~J0901$-$4046 on 20 occasions spanning about 20 hours using the Murriyang radio telescope. The observations were carried out with the ultra-wideband low-frequency (UWL) receiver, which observes over a continuous frequency band covering 704--4032~MHz, divided into 26 individually amplified 128~MHz sub-bands \citep[][]{2020hobbs}. The sub-band voltage data were processed using the `Medusa' GPU-based signal processor system in the pulsar searching mode. In this mode, the \textsc{dspsr} \citep[][]{2011vanstraten} package was used to produce \textsc{psrfits} format data files. The data have an 8-bit resolution, 512~$\upmu$s sampling time, and 3328 frequency channels. The data were de-dispersed to 52.6~pc~cm$^{-3}$ and folded using the timing solution from \citetalias{2022caleb} and \textsc{dspsr}.

The folded data were cleaned of RFI using \texttt{clfd}. Subsequently, affected frequency channels were manually excised using \texttt{pazi}. However, some data were still heavily affected by zero-DM RFI that was difficult to remove due to the low DM of the pulsar. The cleaned Murriyang data were used to produce an average pulse profile over a wide frequency band (see $\S\ref{subsec:pwidth}$) and for pulsar timing purposes (see $\S\ref{subsec:timing}$).

\subsection{GMRT data}\label{subsec:gmrtobs}
A total of $\sim$7~hours worth of observations of PSR~J0901$-$4046 using the GMRT \citep[][]{1991swarup} were carried out across four occasions. They were performed using the upgraded receiver and backend instrumentation \citep[uGMRT;][]{2017gupta}. The observations were carried out in a multi-band phased array configuration in Band-3 (300 to 500~MHz) and Band-4 (550 to 750~MHz). The 8-bit data were collected with 327.68~$\upmu$s sampling time, coherently de-dispersed and divided into 4096 frequency channels in each sub-band using the GMRT Wide-band Backend \citep[GWB;][]{2017reddy}. The data were folded using the PSR~J0901$-$4046 ephemeris and cleaned using \texttt{pazi}. Eight pulses in total were recovered from across all the Band-4 data, but no corresponding Band-3 detections could be identified. The Band-4 detections are included in the time-averaged dynamic spectrum analysis in $\S\ref{subsec:pwidth}$.

\subsection{MWA data}\label{subsec:mwaobs}
On 12 March 2022, a simultaneous two-hour observation targeting PSR~J0901$-$4046 was carried out using MeerKAT and MWA \citep[][]{2013tingay,2018wayth}. The MWA observations were made in the 140$-$170~MHz range. For the first 12 minutes of the observation, only 65 of 128 MWA `tiles' were operational, while for the remaining 108 minutes, 106 tiles were operational. The MWA data were recorded using the Voltage Capture System  \citep[VCS;][]{2015tremblay,2023morrison}{}, which records Nyquist-sampled dual-polarization voltages, from each of the 128 tiles (and hence an aggregate data rate of 56 TB\,h$^{-1}$).  In this mode, the channelized voltages (over 24 $\times$ 1.28-MHz wide coarse channels) are recorded from each tile, which can be processed offline using the software tied-array beamformer \citep{2019ord}. 

The resulting data were processed using the {\tt dspsr} package, and folded using the best available ephemeris, cleaned using \texttt{pazi}, and manually inspected for pulses. However, no PSR~J0901$-$4046 pulses were detected in the data.


\section{Results}\label{sec:results}

\begin{table}
\caption[PSR~J0901$-$4046 timing and model parameters.]{Pulsar timing and model parameters for PSR~J0901$-$4046. Uncertainties in parentheses are 1-$\sigma$ errors on the final significant digit. Note that the listed coordinates are those derived from imaging, rather than from timing.}\label{tab:mtp13_timingpars}
\begin{tabular}{ll}
\toprule
\multicolumn{2}{c}{Data and model fit quality} \\
\midrule
Pulsar name\dotfill & J0901$-$4046 \\ 
MJD range\dotfill & 59119.1---60083.4 \\ 
Data span (yr)\dotfill & 2.60 \\ 
Number of TOAs\dotfill & 55 \\
$n_{\mathrm{dof}}$\dotfill & 52 \\
$\chi^{2}/n_{\mathrm{dof}}$\dotfill &  805.29 \\
RMS timing residual (ms)\dotfill & 7.6 \\
\midrule
\multicolumn{2}{c}{Measured quantities} \\ 
\midrule
Right ascension, $\alpha$ (hh:mm:ss)\dotfill & 09$^h$01$^m$29.249$^s$ $\pm \ang{;;1.0}$\\ 
Declination, $\delta$ (dd:mm:ss)\dotfill & $\ang{-40;46;02.984} \pm \ang{;;1.0}$\\ 
Pulse period, $P$ (s)\dotfill & $ 75.88554698  \pm 3\times10^{-8}$\\
First derivative of pulse period, $\dot{P}$ (s~s$^{-1}$)\dotfill & $\left(2.44 \pm 0.01 \right) \times 10^{-13}$ \\
Dispersion measure, DM (cm$^{-3}$pc)\dotfill & $52.6\pm1$ \\ 
\midrule
\multicolumn{2}{c}{Inferred quantities} \\ 
\midrule
Distance (\textsc{ymw16}), $d_{1}$ (pc)\dotfill & 328\\
Distance (\textsc{ne2001}), $d_{2}$ (pc)\dotfill & 467\\
Characteristic age, $\tau$ (Myr)\dotfill & 4.9\\
Surface dipole magnetic field strength, $B$ (G)\dotfill & 1.4$\times10^{14}$\\
Spin-down luminosity, $\dot{E}$ (erg s$^{-1}$)\dotfill & 2.2$\times10^{28}$\\
\bottomrule
\end{tabular}
\end{table}

\subsection{Timing}\label{subsec:timing}

The MeerKAT FBFUSE data ($\S\ref{subsec:meerkatobs}$), Murriyang UWL data ($S\ref{subsec:pksobs}$), and GMRT Band-3 and Band-4 data ($\S\ref{subsec:gmrtobs}$) were used to derive Times of Arrival (ToAs) for PSR~J0901$-$4046 single pulses. These were used in conjunction with the MeerKAT L band and Murriyang UWL ToAs reported in \citetalias{2022caleb} to calculate a coherent timing solution for the pulsar.

\begin{figure*}
    \centering
    \includegraphics[width=\textwidth]{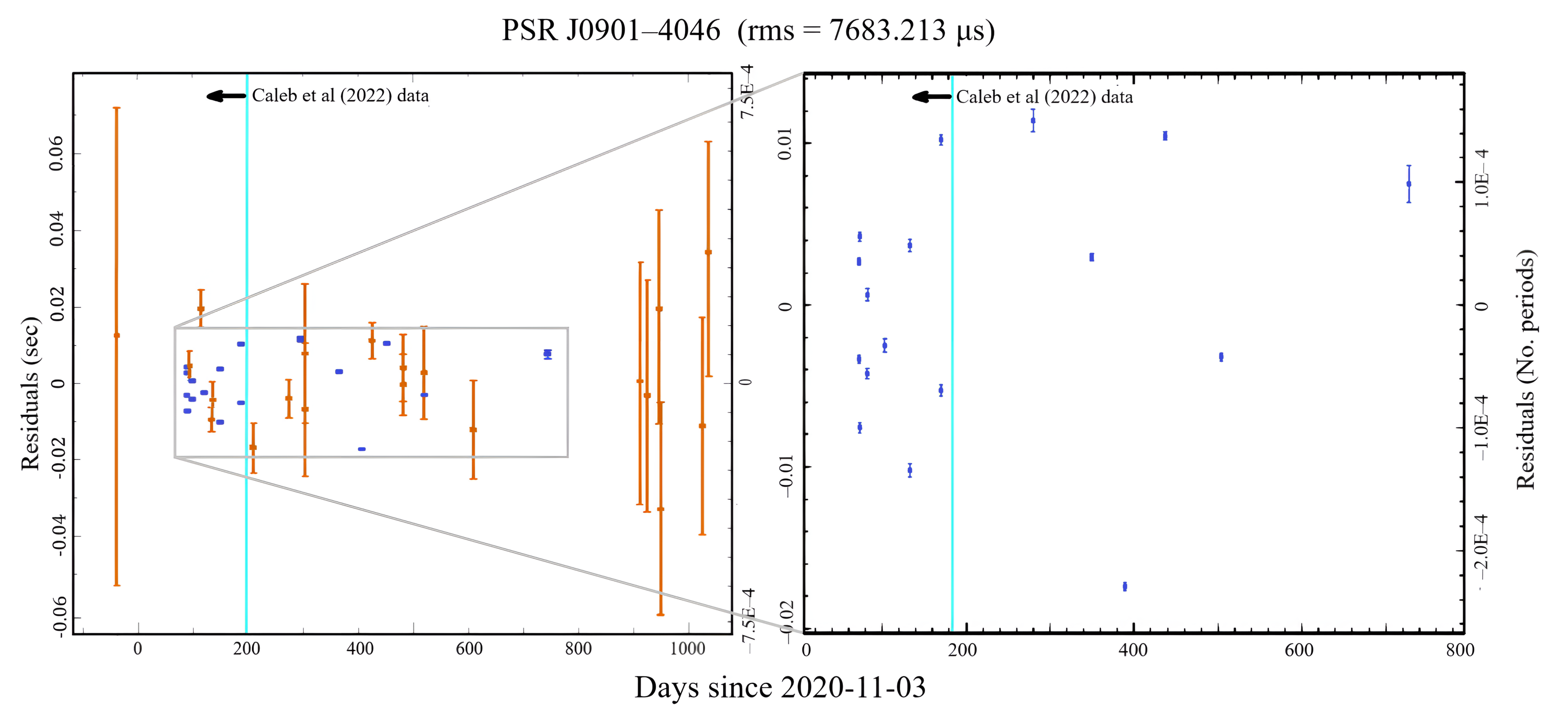}
    \caption[PSR~J0901$-$4046 timing residuals.]{Timing residuals of PSR~J0901$-$4046 pulses. Blue data points are determined from MeerKAT data, while orange data points are determined from Murriyang UWL data. The right-hand panel is zoomed in on the MeerKAT data points only. The vertical cyan line separates the data points used for the timing analysis in \citetalias{2022caleb} from subsequent measurements. The error bars are 1-$\sigma$.}
    \label{fig:mtp13_timing}
\end{figure*}

The \textsc{psrchive} pulsar analysis library was used to process the data. For each set of observations, we used the de-dispersed and cleaned single pulse archives to produce an averaged pulse profile. The \textsc{psrchive} tool \texttt{paas} was then used to create an analytic template profile for each data set. We used \texttt{pat} to generate ToAs for all pulses. Timing was then performed using the \textsc{tempo2} pulsar timing software \citep[][]{2006hobbs}. In addition to the pulse period $P$ and period derivative $\dot{P}$, a jump between the UWL data and MeerKAT data was included in the timing model. The timing parameters from the resulting coherent timing solution are listed in Table~\ref{tab:mtp13_timingpars}. Note that we do not fit for position, since the uncertainty on the resulting fit is larger than the $\ang{;;1}$ uncertainty on the imaging-derived source position from \citetalias{2022caleb}, 09:01:29.25(7), $-$40:46:03(1). If RA and Dec are included in the fit, the best-fit coordinates are 09:01:29.2(1), $-$40:46:02(2), and the timing residual RMS is 7.5~ms.

The residuals from the derived timing solution are shown in Fig.~\ref{fig:mtp13_timing}. Some pulse phase jitter is visible where the scatter in arrival times is larger than the error bars. However, the overall timing RMS is proportionally very small, being roughly 1/10000th of the pulse period at 7.6~ms. This makes the pulsar exceptionally well-timed, and suggests that the shape of the pulse profile envelope is highly stable from epoch to epoch. For comparison, typical millisecond pulsars have timing residuals with RMS of around 0.1--1~$\mu$s, or around 10$^{-5}$--10$^{-3}$ of the pulse period \citep[][]{2018becker}.

The measured $P$ of 75.88554698(3)~s is consistent with the value of 75.88554711(6) reported in \citetalias{2022caleb}, while the measured $\dot{P}$ of 2.44(1)$\times10^{-13}$~s~s$^{-1}$ agrees with the value in \citetalias{2022caleb} of 2.25(10)$\times10^{-13}$~s~s$^{-1}$ within 2-$\sigma$. These timing parameters imply a characteristic magnetic field strength of $B = 3.2\times10^{19}\sqrt{P\dot{P}}\simeq1.4\times10^{14}$~G, a characteristic age of $\tau = P/2\dot{P} = 4.9$~Myr, and a spin-down energy of $\dot{E} = 4\pi^{2}I\dot{P}/P^{3} = 2.2\times10^{28}$~erg s$^{-1}$.

\subsection{Average profile and pulse width evolution}\label{subsec:pwidth}
\begin{figure}
    \centering
    \includegraphics[width=.49\textwidth]{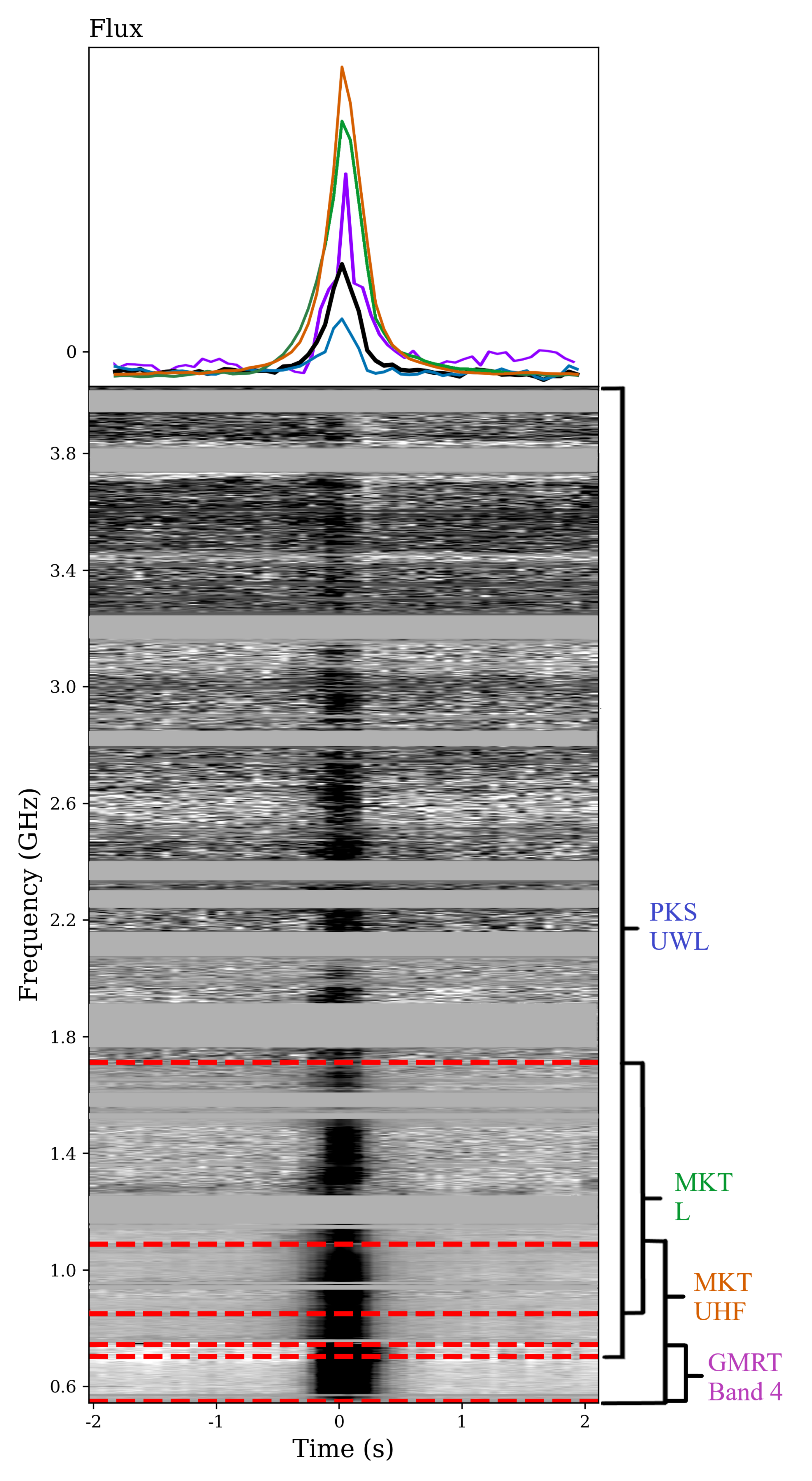}
    \caption[Combined Murriyang UWL, MeerKAT L band and UHF band, and GMRT Band-4 profile and dynamic spectrum of PSR~J0901$-$4046.]{Profile (top) and dynamic spectrum (bottom) of PSR~J0901$-$4046 averaged over the Murriyang UWL band, GMRT Band-4, and the MeerKAT L band and UHF band. The fluxes are indicated in arbitrary units. Profiles averaged over the combined band (black), UWL only (blue), Band-4 only (violet), MeerKAT L band only (green), and MeerKAT UHF band only (orange) are shown. The red horizontal lines in the dynamic spectrum indicate the regions where the respective bands overlap. The solid grey regions correspond to channels removed due to persistent RFI.}
    \label{fig:mtp13_fbfuse+pks_sum}
\end{figure}

The Murriyang UWL, MeerKAT FBFUSE, and GMRT data were used to produce an average profile across a wide bandwidth from 544 to 4032~MHz, which is presented in Fig.~\ref{fig:mtp13_fbfuse+pks_sum}. After being cleaned of RFI, the data were scrunched using the \textsc{psrchive} tool \texttt{pam} to a single polarisation channel with 1024 phase bins and 1024 frequency channels. The UWL and MeerKAT L-band data were then downsampled so that their frequency channel width matches that of the UHF band data. The data were also scaled so that they have the same root-mean-square (RMS) in each band. Where the respective frequency bands overlap, the data were averaged. It is evident that PSR~J0901$-$4046 is visible over a wide range of frequencies with little discernible frequency structure. 

The integrated profile of PSR~J0901$-$4046 has a half-power pulse width of $\sim$300~ms at $\sim 1.2$~GHz, resulting in an extremely narrow duty cycle of $\sim1.4^{\circ}$. This is more in line with other radio pulsars with $P>5$~s than with the known radio-loud magnetars, which have duty cycles $\sim$15--190$^{\circ}$ \citep[see e.g.][]{2021agar}. Moreover, this width appears to be stable over a wide bandwidth, with similar values observed at the bottom of the MeerKAT UHF band at 544~MHz and the top of the Murriyang UWL band at 4032~MHz.

The width of pulse shapes of non-recycled pulsars is most often observed to vary monotonically across the observing band, but most prominently below 1~GHz, a behaviour commonly associated with the putative radius-to-frequency mapping (RFM) mechanism \citep[][]{1991thorsett}. The model proposed by \citet{1978cordes} stipulates that lower frequency radiation is emitted further from the NS surface, where the open magnetic field lines possess a wider opening angle, resulting in a beam that widens towards low frequencies.

\citet[][]{1991thorsett}  modelled pulsar profile width evolution using the following relation:
\begin{equation}
    W_{10} = A\nu^{\mu} + W_{10,0},
    \label{eq:thorsett}
\end{equation}
where W$_{10}$ is the profile width at 10~per~cent of the maximum, $\nu$ is the observing frequency in MHz, A and $\mu$ are constants, and $W_{10,0}$ is the asymptotic profile width.

\citet{1998kijak} evaluated the pulse width of 87 pulsars at 1.4~GHz and 4.85~GHz, finding that only 57~per~cent of the pulsars were more than 5~per~cent narrower at the higher frequency compared to the lower frequency; 41~per~cent of the pulsars exhibited marginal variability, while 2~per~cent were significantly wider at the higher frequency. \citet{2008johnston}  found that the profiles of around 30 per cent of 67 pulsars observed at 243 to 3100~MHz broadened at higher frequencies. Similarly, applying Eq.~\ref{eq:thorsett} to 150 pulsars in the European Pulsar Network demonstrated that 54~per~cent narrowed towards higher frequencies, 27~per~cent were approximately constant in width, and 19~per~cent grew wider \citep{2014chen}. Subsequently, \citet{2016pilia}, \citet{2019olszanski}, \citet{2019zhao}, and \citet{2021posselt} have found similar frequency dependence of pulse width among pulsars observed with the LOFAR, Arecibo, GMRT, and MeerKAT telescopes, respectively. Frequency-dependence is also seen in pulsars that exhibit phase modulation phenomena such as e.g. sub-pulse drifting, where it is generally seen as an increase in the component separation at lower frequencies \citep[e.g.][]{2023mcsweeney}.

To study the RFM-like behaviour of PSR~J0901$-$4046, we used the wideband, composite dynamic spectrum shown in Figure~\ref{fig:mtp13_fbfuse+pks_sum}. The average pulse was divided into 20 frequency channels of differing widths, such that S/N $> 10$ in each.  For each channel, the baseline was subtracted using the \textsc{psrsalsa}\footnote{\url{https://github.com/weltevrede/psrsalsa}} software package \citep[][]{2016weltevrede}, and a pulse profile template was derived using \textsc{psrchive}/\texttt{paas}. The \texttt{fitvonMises} function in \texttt{PSRSALSA} was then used to measure the best-fit W$_{10}$ value and error. This was repeated for each channel to obtain width measurements across the frequency band. We then used the \texttt{lmfit} Python package \citep{2014newville} to carry out a least-square minimisation fit of Eq.~\ref{eq:thorsett} to the measured widths. The pulse widths (in degrees of the phase) are shown in Figure~\ref{fig:pulse_widths}, along with the best-fit Thorsett function. Best-fit parameters of $\mathrm{A} = 5300\pm2700$, $\mu=-2.14\pm0.47$, and $\mathrm{W}_{10,0} = 1.98^{\circ}\pm0.09$ were obtained. 

\citet{2021posselt} demonstrated that the power-law parameters are strongly correlated, and therefore only weakly determinative of frequency-dependent behaviour when considered individually. The extrapolated fractional width change from 1.4~GHz to 4.85~GHz given by 
\begin{equation}
    \eta\prime = \frac{W_{4.85 \mathrm{GHz}} - W_{1.4 \mathrm{GHz}}}{W_{1.4 \mathrm{GHz}}},
    \label{eq:eta_prime}
\end{equation}  
from \citet{2014chen} is a more indicative point of comparison. The best-fit parameters listed above return a value of $\eta\prime = 0.00 \pm 0.02$ for PSR~J0901$-$4046, indicating a marginal, if any, change in the pulse width as a function of frequency. In Figure~\ref{fig:width_histograms}, histograms of the best-fit $\mu$ and $\eta\prime$ values of the 150 pulsars studied by \citet{2014chen} are shown, with the derived values for PSR~J0901$-$4046 overlaid.

The best-fit power-law index $\mu$ is significantly shallower than what would be expected for Kolmogorov turbulence scattering, i.e.~$\sim-4.4$, indicating weak presence of scattering. Assuming that the pulse profile is well-modelled by a Gaussian function, the full width at half maximum of the pulse should be given by 
\begin{equation}
    W_{50} = \sqrt{\frac{\ln 2}{\ln10}} W_{10}.
\end{equation}
 Then, the obtained for fit Eq.\ \ref{eq:thorsett} provides an upper limit on the scattering timescale in the case that scattering dominates: 
 \begin{equation}
     \tau_{\mathrm{sc}}\left( \nu \right) < W_{50}(\nu) = \sqrt{\frac{\ln2}{\ln 10}} (A \nu ^{\mu} + W_{10,0}),
 \end{equation}
so that $\tau_{\mathrm{sc,1GHz}} \lesssim 250$~ms. By comparison, the $\textsc{NE2001}$ \citep[][]{ne2001} and $\textsc{YWM16}$ \citep[][]{ymw16} Galactic electron distribution models predict scattering timescales along the pulsar's line of sight at a DM of 52.6 pc\ cm$^{3}$ of 1.6~$\mu s$ and 7.0~ns, respectively.

\subsection{Pulse energies}
\label{sec:penergy}
\subsubsection{Average spectrum}
\label{subsec:spectrum}
Using the time-averaged pulse data shown in the previous section, we obtained spectral fits for the pulsar in each sub-band. We fit the power-law function
\begin{equation}
    S(\nu) = A\nu^{\alpha}
\end{equation}
to the measured intensities, omitting data points that were greater than 3-$\sigma$ away from the median value. The observed intensity values are well-fit by the simple power law, with no evidence of variation as a function of frequency observed. We obtained best-fit spectral indices of $\alpha= -0.92\pm0.01$, $-1.15\pm0.12$, and $-1.02\pm0.05$ across the MeerKAT UHF band, MeerKAT L band, and Murriyang UWL band, respectively.

\begin{figure}
    \centering
    \includegraphics[width=\linewidth]{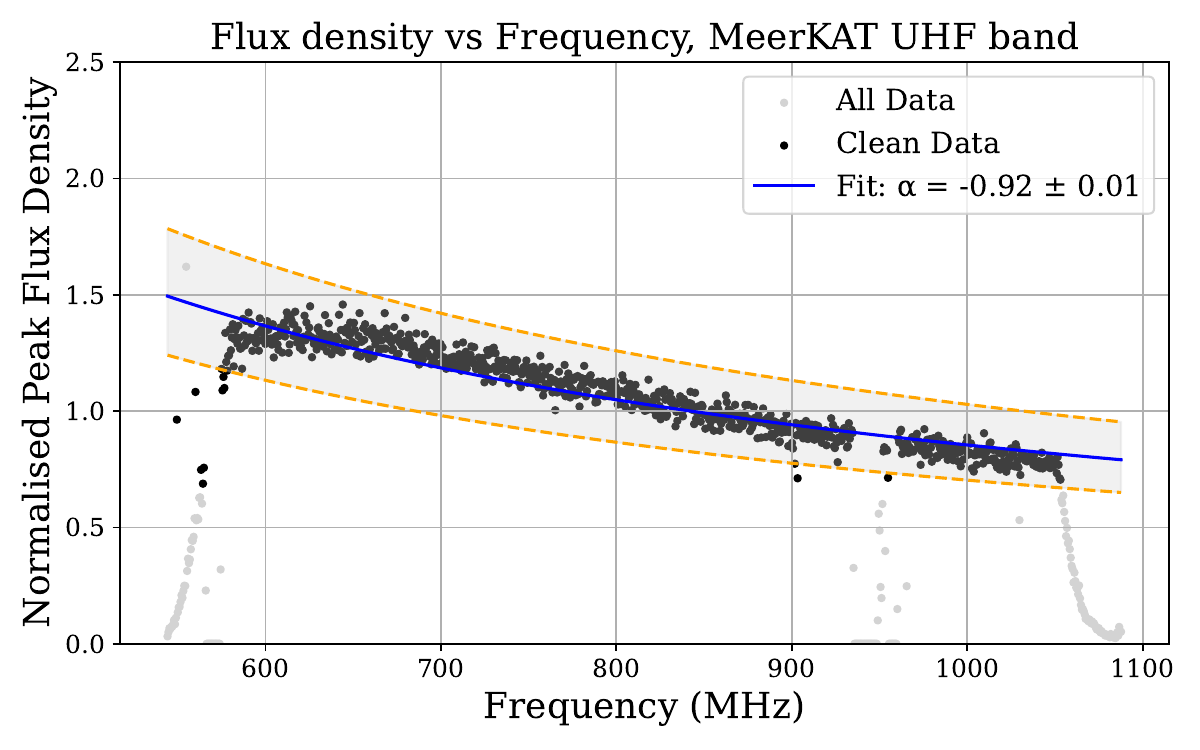}
    \caption{Measured time-averaged pulse energies as a function of frequency in the MeerKAT UHF band, with best-fit power law spectral fit. The gray data points were ignored in the fit as outliers.}
    \label{fig:enter-label}
\end{figure}

\begin{figure}
    \centering
    \includegraphics[width=.5\textwidth]{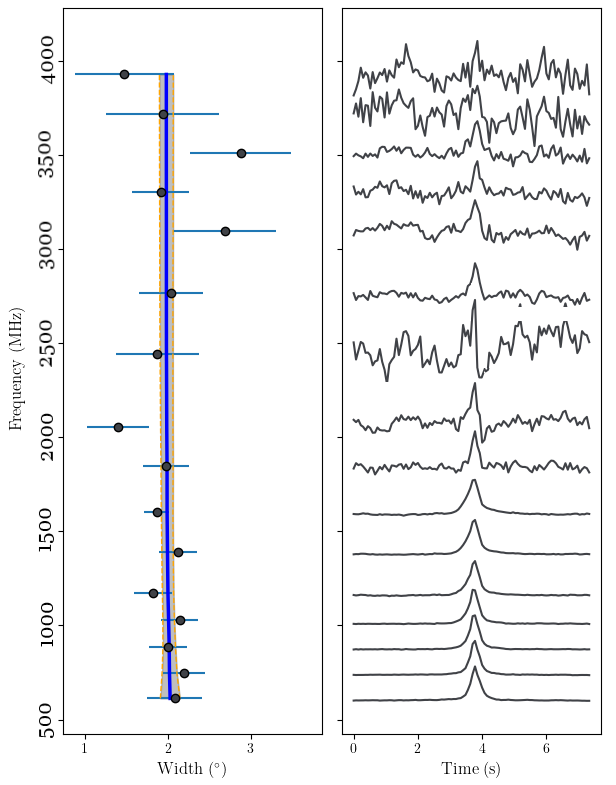}
    \caption{Left: W$_{10}$ pulse widths of PSR~J0901$-$4046 in 20 frequency channels covering 0.5--4~GHz. The blue line represents the best-fit model using Eq.~\ref{eq:thorsett}, with 1-$\sigma$ errors shaded. Right: The pulse profile in each channel.}
    \label{fig:pulse_widths}
\end{figure}

\subsubsection*{Pulse energy distribution}
We used single pulse archives from the seven-hour PTUSE observation on 26 September 2021 to obtain pulse energies. To do this, \textsc{psrsalsa} was used to first flatten the slope of the baseline in the data, which is accentuated by the very long period, as well as to gate the data to include only 130 phase bins around the pulse peak. For some pulses, zero-DM RFI spikes nearby the pulse peak had to be removed by hand using \texttt{pazi}. The gating and zero-DM RFI excision of the data aided in the subsequent subtraction of the baseline from the data, also using \textsc{psrsalsa}, which was complicated both by the presence of RFI with DM close to the pulsar DM and the extreme width of the full pulse period. 151 single pulses were identified in these data that were detected in both the L band and UHF band concurrently.

\begin{figure}
    \centering
    \includegraphics[width=.5\textwidth]{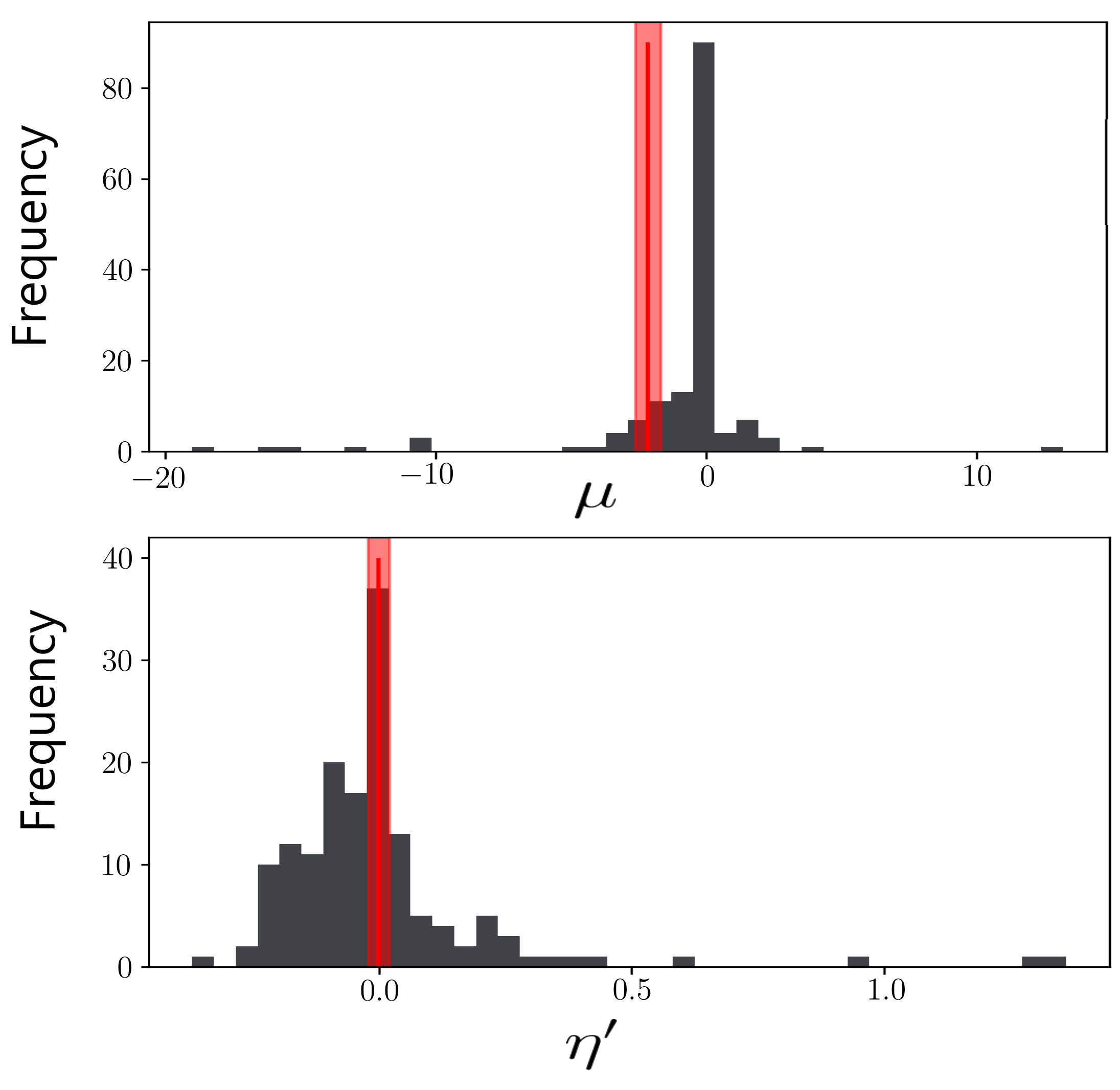}
   \caption{Histograms comparing the best-fit $\mu$ (top) and $\eta\prime$ (bottom) parameters with shaded errors for PSR~J0901$-$4046 compared to the 150 pulsars in the \citet{2014chen} sample.}
    \label{fig:width_histograms}
\end{figure}

\begin{figure*}
    \centering
    \includegraphics[width=\textwidth]{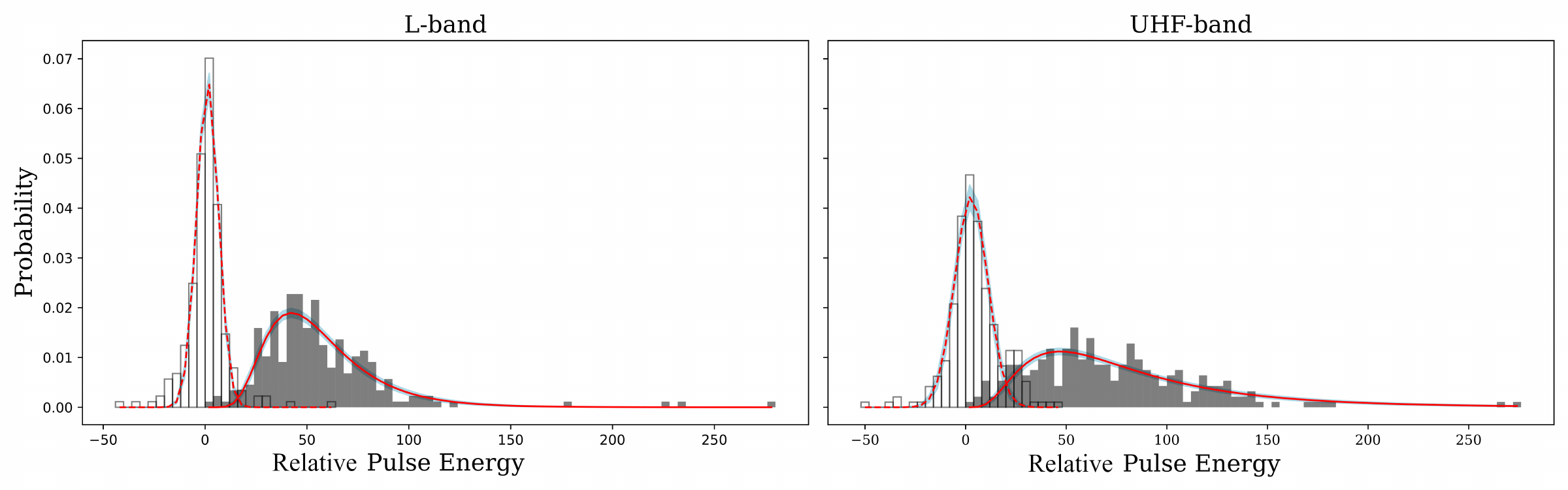}
    \caption[PSR~J0901$-$4046 single pulse energy distribution in the MeerKAT L band and UHF band.]{Pulse energy distribution of PSR~J0901$-$4046 single pulses observed simultaneously in the MeerKAT L band (left) and UHF band (right). In each plot, the black outlined bars indicate the off-pulse energy, while the solid grey bars indicate the on-pulse energy. The energies are normalised with respect to the average off-pulse energy. The red curves show best-fit log-normal functions fit to the off-pulse (dashed) and on-pulse (solid) energies, with 1-$\sigma$ errors in cyan.}
    \label{fig:mtp13_ped}
\end{figure*}

In order to compare the distribution of pulse energies in the different MeerKAT bands, which have different noise characteristics, we computed relative pulse energies in both bands for the 151 identified pulses using \textsc{psrsalsa}. This involved identifying an on-pulse and off-pulse region of each pulse's profile, measuring the on-pulse and off-pulse flux, and then normalising with respect to the average off-pulse value. While the on-pulse phase range was specified to be the same for all single pulses, the off-pulse region was manually chosen for each pulse to include the same number of phase bins as the on-pulse region, but not to coincide with any phase bins where a zero-DM RFI spike had been removed. The on-pulse and off-pulse energy distributions (PEDs) derived for all the PTUSE detections of PSR~J0901$-$4046 in each band are shown in Fig.~\ref{fig:mtp13_ped}. For each distribution, we fit a Gaussian function,
\begin{equation}\label{eq:gaussian}
    f(E) = \frac{A}{\sigma \sqrt{2\pi}}\exp{\left(\frac{-\left(E-\mu\right)^{2}}{2\sigma^{2}}\right)}
\end{equation}
to the off-pulse energies, and a log-normal function,
\begin{equation}\label{eq:lognormal}
    f(E) = \frac{A}{E\sigma\sqrt{2\pi}} \exp{\left( \frac{-\left(\ln\left(E\right)-\mu\right)^{2}}{2\sigma^{2}} \right)},    
\end{equation}
to the on-pulse energies. The best-fit $A$, $\mu$, and $\sigma$ parameters, as well as reduced $\chi^{2}$ values, for each fit are recorded in Table \ref{table:MTP13_PED_fits}. Note that the combination of RFI, variability across the pulse, and the variable baseline makes determining the pulse energies of weaker pulses difficult, and also broadens the off-pulse energy distribution.

The off-pulse energy distributions are both centred around zero, and comport with the expected Gaussian noise characteristics, while the greater width of the on-pulse distribution in the UHF band confirms the finding in \citetalias{2022caleb} that the pulse-to-pulse modulation of the source is greater at lower frequencies. 

\begin{table*}
\centering
\caption[PSR~J0901$-$4046 pulse energy distribution best-fit parameters and reduced $\chi^{2}$ values.]{The best-fit parameters and $\chi^{2}$ values for Gaussian and log-normal functions fit to the on-pulse and off-pulse energy distributions shown in Figure~\ref{fig:mtp13_ped}, respectively}
\label{table:MTP13_PED_fits}
\begin{tabular}{lrrrrrrrr}
\toprule
& \multicolumn{4}{c}{Off-pulse energies} & \multicolumn{4}{c}{On-pulse energies}\\
\cmidrule(lr){2-5}\cmidrule(lr){6-9}
Frequency band & A & $\mu$ & $\sigma$ & $\chi^{2}_r$ & A & $\mu$ & $\sigma$ & $\chi^{2}_r$\\
\midrule
UHF-band & 0.87$\pm$0.05 & 2.8$\pm$0.5 & 8.2$\pm$0.5 & 1.4e$-$4 & 1.03$\pm$0.07 & 4.26$\pm$0.05 & 0.64$\pm$0.04 & 3.8e$-$6\\
L-band & 0.87$\pm$0.03 & 1.2$\pm$0.2 & 5.3$\pm$0.2 & 8.2e$-$6 & 0.98$\pm$0.05 & 3.94$\pm$0.03 & 0.44$\pm$0.02 & 4.7e$-$6\\
\bottomrule
\end{tabular}
\end{table*}

\subsubsection*{Temporal variability}
For each pulse detected with MeerKAT, we used the S/N-DM optimisation code \texttt{mtcutils}\footnote{\url{bitbucket.org/vmorello/mtcutils}} to calculate the peak S/N. These values were then converted to estimated peak flux density $S_{\nu}$ using the radiometer equation for single pulses \citep[][]{2012lorimer}:
\begin{equation}
    S_{\nu} \simeq \frac{S/N~T_{\mathrm{sys}}}{G~f_\mathrm{ant}~\sqrt{n_\mathrm{p}~W~\Delta f}},
\end{equation}
where $G$ is the telescope's gain, $T_{\mathrm{sys}}$ is the system temperature, $f_\mathrm{ant}$ is the fraction of antennas used in the observation, $n_\mathrm{p}$ is the number of polarisation channels recorded, $W$ is the pulse width, and $\Delta f$ is the bandwidth. The system temperature is assumed to be the sum of the receiver temperature and the sky temperature for the observation, $T_{\mathrm{sys}} =T_{\mathrm{rec}} + T_{\mathrm{sky}}$. We used instrumental parameters of $T_{\mathrm{rec}} = 18$~K at L-band and 24~K at UHF-band, $G= 2.8$~K/Jy (adjusted for the number of dishes in each observation), $n_\mathrm{p} = 1$, and $\Delta f = 856$~MHz at L-band and 544~MHz at UHF-band \citep[][]{2020bailes}. The pulse widths were measured with \texttt{mtcutils}. The $T_{\mathrm{sky}}$ values at the source coordinates at the centre of the centre of the observing band were determined using the \texttt{pygdsm} \citep[][]{2016price} implementation of the diffuse Galactic radio emission model of \citet{2017zheng}.

\begin{figure*}
    \centering
    \includegraphics[width=\textwidth]{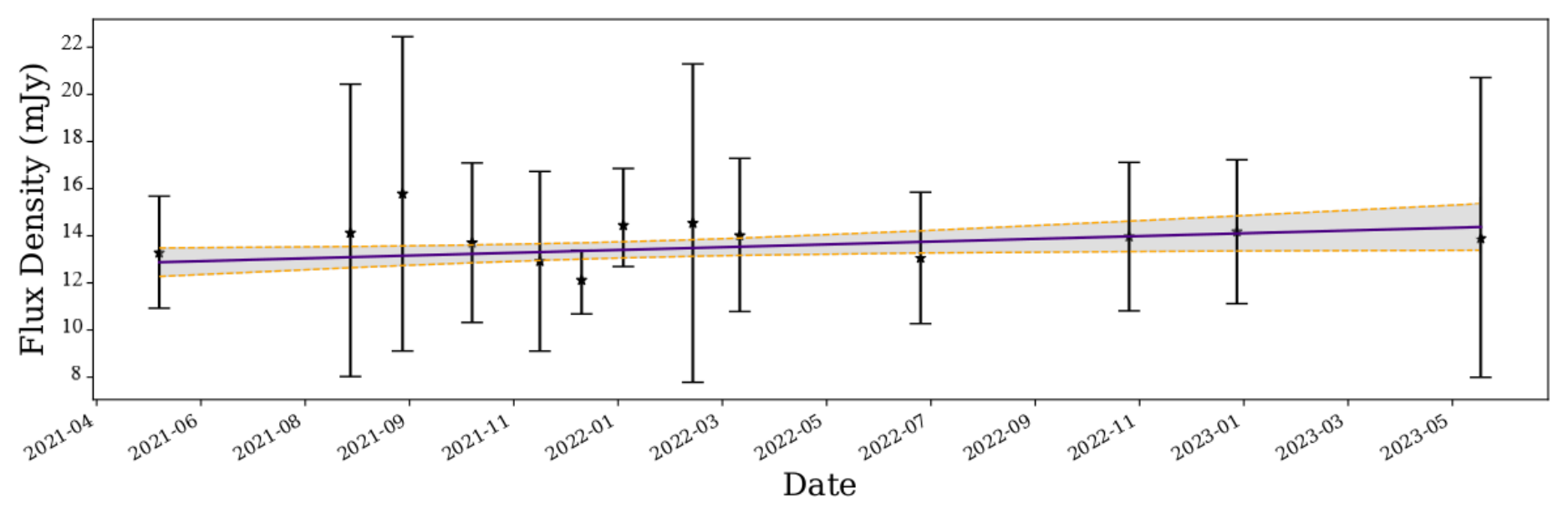}
    \caption[Average peak flux density evolution of PSR J0901$-$4046]{Average peak flux density measurements of single pulses detected during MeerKAT observations. The line of best fit, with 1-$\sigma$ uncertainty, has a slope of 4.8$\pm$2.1$\times$10$^{-3}$ mJy/day.}
    \label{fig:flux_evol}
\end{figure*}

The radiometer equation-derived flux density values measured across all MeerKAT observations of the source are shown in Figure~\ref{fig:flux_evol}. A weighted linear regression using \texttt{lmfit} returned a best-fit line with slope 4.8$\pm$2.1$\times$10$^{-3}$ mJy/day. We therefore do not observe the possible trend of secular decline in the pulse brightness identified in \citep[][]{2022caleb}. Instead, the pulses grew marginally brighter over time, likely indicative of short-term variability more than any long-term trend.

\begin{figure}
    \centering
    \includegraphics[width=.48\textwidth]{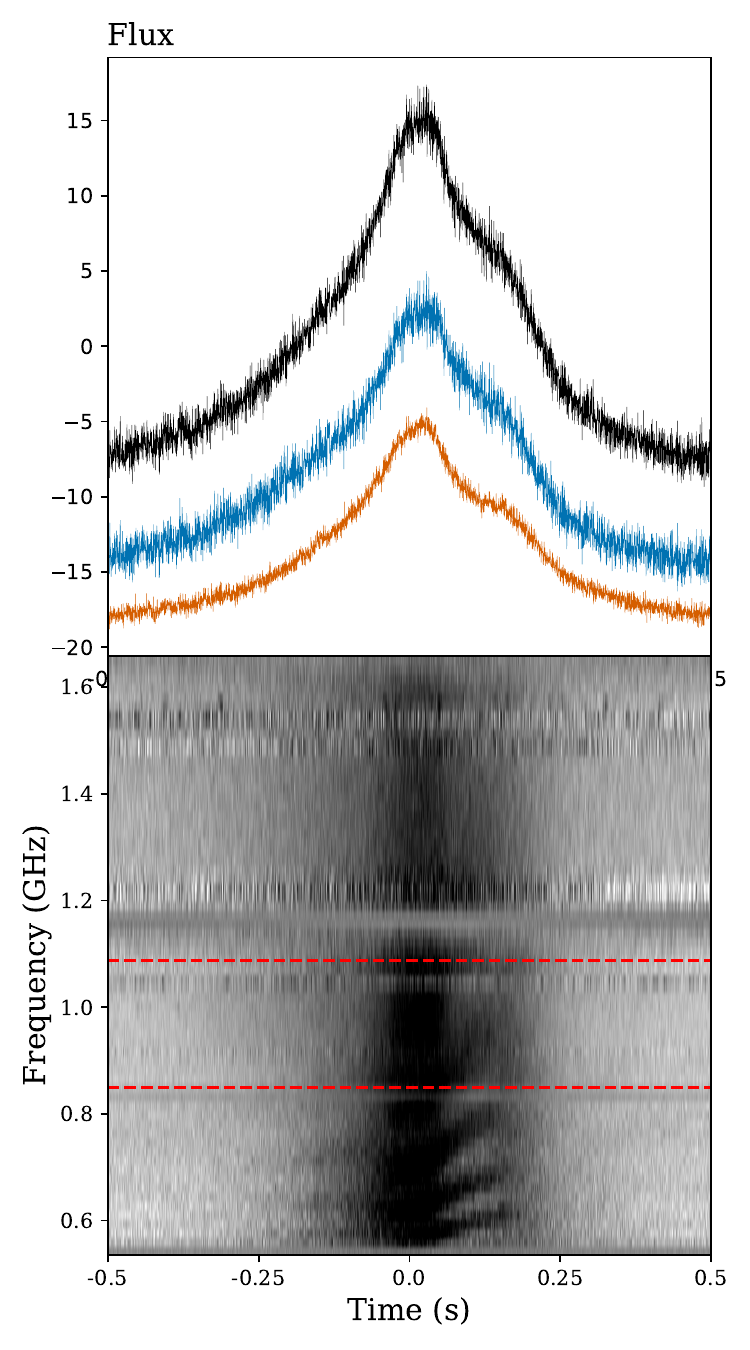}
    \caption[High time-resolution MeerKAT profile and dynamic spectrum of PSR J0901$-$4046]{High time-resolution profile (top) and dynamic spectrum (bottom) made using 151 PSR~J0901$-$4046 single pulses observed in the MeerKAT L band and UHF band concurrently. Profiles summed over the combined band (black), MeerKAT L-band only (blue), and MeerKAT UHF-band only (orange) are shown. The red horizontal lines in the dynamic spectrum indicate the regions where the respective bands overlap. The higher noise regions near 1.2~GHz and 1.5~GHz are affected by satellite RFI. The apparent striations at low frequencies are caused by the removal of zero-DM RFI.}
    \label{fig:mtp13_ptuse_sum}
\end{figure}

\subsection{Pulse shapes}
\label{subsec:pmorph}

\citetalias{2022caleb} noted certain remarkable spectro-temporal-polarimetric properties of PSR~J0901$-$4046's single pulses. In particular, the observed pulse morphologies varied drastically from one instance to the next. \citetalias{2022caleb} identified seven distinct categories of PSR~J0901$-$4046 pulse morphology:~\textit{normal}, \textit{double peaked}, \textit{triple peaked}, \textit{split peaked}, \textit{quasi-periodic}, \textit{partially nulling}, and \textit{spiky} (see Fig.~2 of that paper for an example of each). Unlike the sub-pulse structure seen in pulses from e.g.~some radio-loud magnetars, the pulse type variation seemed chaotic, with no discernible pattern. Additionally, certain pulses exhibited one or multiple quasi-periodic sub-pulse components, which is behaviour reminiscent of some FRBs \citep[e.g.][]{2021chime} as well as the radio-loud magnetar XTE~J1810$-$197 \citep[][]{2019levin,2022caleb2}. 

The PTUSE data from the 26 September 2021 dual-band MeerKAT observation were used to study PSR~J0901$-$4046's pulse shape and sub-pulse behaviour. For this purpose, the data were integrated to a single polarisation channel with 65536 phase bins and 32 frequency channels. The data were also gated, and the baseline removed using \textsc{psrsalsa} as outlined in $\S\ref{sec:penergy}$.

\begin{figure*}
    \centering
    \includegraphics[width=\textwidth]{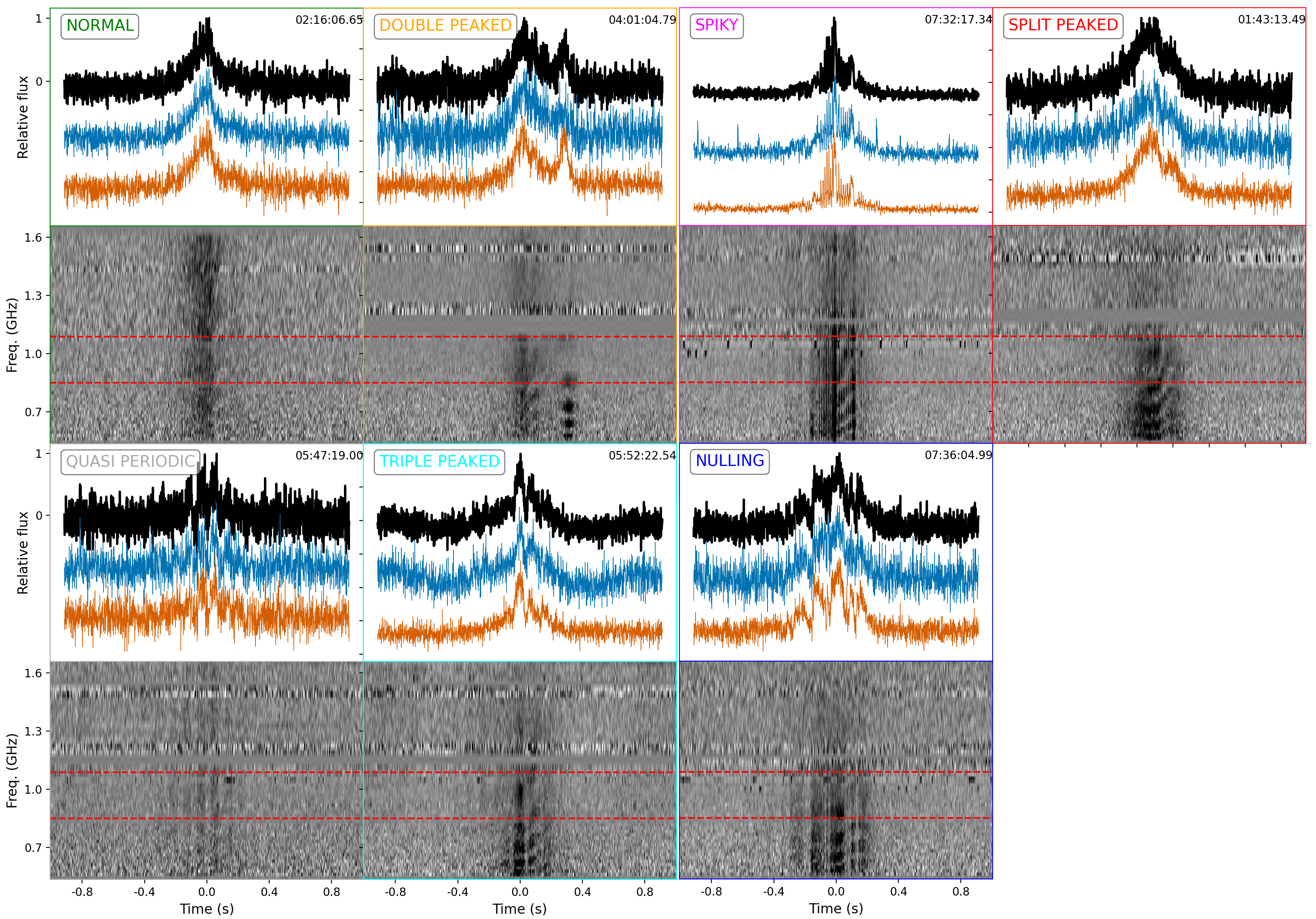}
    \caption[Representative PSR~J0901$-$4046 single pulses of each pulse morphology type.]{Profiles (top) and dynamic spectra (bottom) of representative single pulses from each of the seven PSR~J0901$-$4046 pulse morphology types identified in \citetalias{2022caleb}. The profile averaged over the combined band (black), L band only (orange), and UHF band only (blue) are shown. The red horizontal lines in the dynamic spectra indicate the region where the L band and UHF band overlap. The apparent striations at low frequencies are caused by the removal of zero-DM RFI. The detection time (UTC) for each pulse is listed at the top of each panel.}
    \label{fig:mtp13_shape_examples}
\end{figure*}

Subsequently, pulse profiles were created for each of the 151 PSR~J0901$-$4046 single pulses detected in both the L band and UHF band. The average, high time-resolution pulse profile and dynamic spectrum are shown in Fig.~\ref{fig:mtp13_ptuse_sum}. The profiles were then inspected by eye and classified into one of the seven pulse shape categories listed in \citetalias{2022caleb}.  Fig.~\ref{fig:mtp13_shape_examples} shows representative examples of each of the seven morphological types. All 151 single pulses are plotted in Appendix A of \citet{2022bezuidenhoutThesis}, with their assigned morphological type indicated. It should be borne in mind that, due to the prevalence of zero-DM RFI in the data, the low intensity of some pulses, and the inherently subjective nature of the by-eye classification process, a proportion of the classifications are unavoidably ambiguous.

Intriguingly, the pulse morphologies we identified seem to be distributed differently than reported in \citetalias{2022caleb}. Table~\ref{table:shape_types} compares the number and proportions of pulses of each shape type detected in the two datasets. Although in both cases \textit{normal} pulses were the most common type, they dominated the data analysed in this work, comprising $68\%$ of all identified pulses, while the \citetalias{2022caleb} analysis found a much more even spread of pulse types. Our sample of pulses also contained significantly fewer quasi-periodic and partially nulling pulses (five and three, respectively), despite together comprising 30~per~cent of pulses in \citetalias{2022caleb}. 

\begin{table}
\centering
\caption{Number and proportion of different pulse shapes identified in the \citetalias{2022caleb} sample of pulses and those identified in this work.}
\label{table:shape_types}
\begin{tabular}{lcccc}
\toprule
 & \multicolumn{2}{c}{\citetalias{2022caleb}} & \multicolumn{2}{c}{This work}\\
\cmidrule(lr){2-3}\cmidrule(lr){4-5}
Shape type & Number & Proportion & Number & Proportion\\
\midrule
Normal & 73 & 37\%  & 102 & 68\% \\
Split peaked & 42 & 21\% & 25 & 17\%\\
Quasi-periodic & 31 & 16\% & 5 & 3\% \\
Nulling & 28 & 14\% & 3 & 2\%\\
Spiky & 17 & 9\% & 8 & 5\%\\
Double peaked & 3 & 2\% & 5 & 3\% \\
Triple peaked & 4 & 2\% & 3 & 2\%\\
\midrule
 & 198 &  & 151 & \\
\bottomrule
\end{tabular}
\end{table}

A $\chi^{2}$ test for homogeneity of the observed proportions results in a test statistic of $\chi^{2}=46.9$, exceeding the critical value of 16.8 for six degrees of freedom at the $1\%$ confidence level. This indicates that the pulse shapes observed by \citetalias{2022caleb} are not consistent with the same underlying distribution as those observed in this work, and that the difference is therefore likely not a result of Poissonian variation.

The differing composition of the pulse morphologies observed in the data analysed for this work compared to that seen during the observations spanning February to May 2021 \citepalias{2022caleb} may indicate an evolution of the source. Although no definite pulse-shape trend could be identified in the \citetalias{2022caleb} data, the predominant single pulse shape did seem to vary somewhat from one observing date to the next. For example, of the 24 pulses identified in the MeerKAT L-band observations on 2 April 2021, only three ($\sim$13~per~cent) were of the \textit{split-peak} variety, while \textit{normal} pulses were most common on that occasion, making up $\sim$46~per~cent of all pulses. Further analysis of PSR~J0901$-$4046 single pulse types over time may demonstrate that its emission assumes phases during which one pulse type or another is predominant, depending on the physical state of the source.

\begin{figure*}
    \centering
    \includegraphics[width=\textwidth]{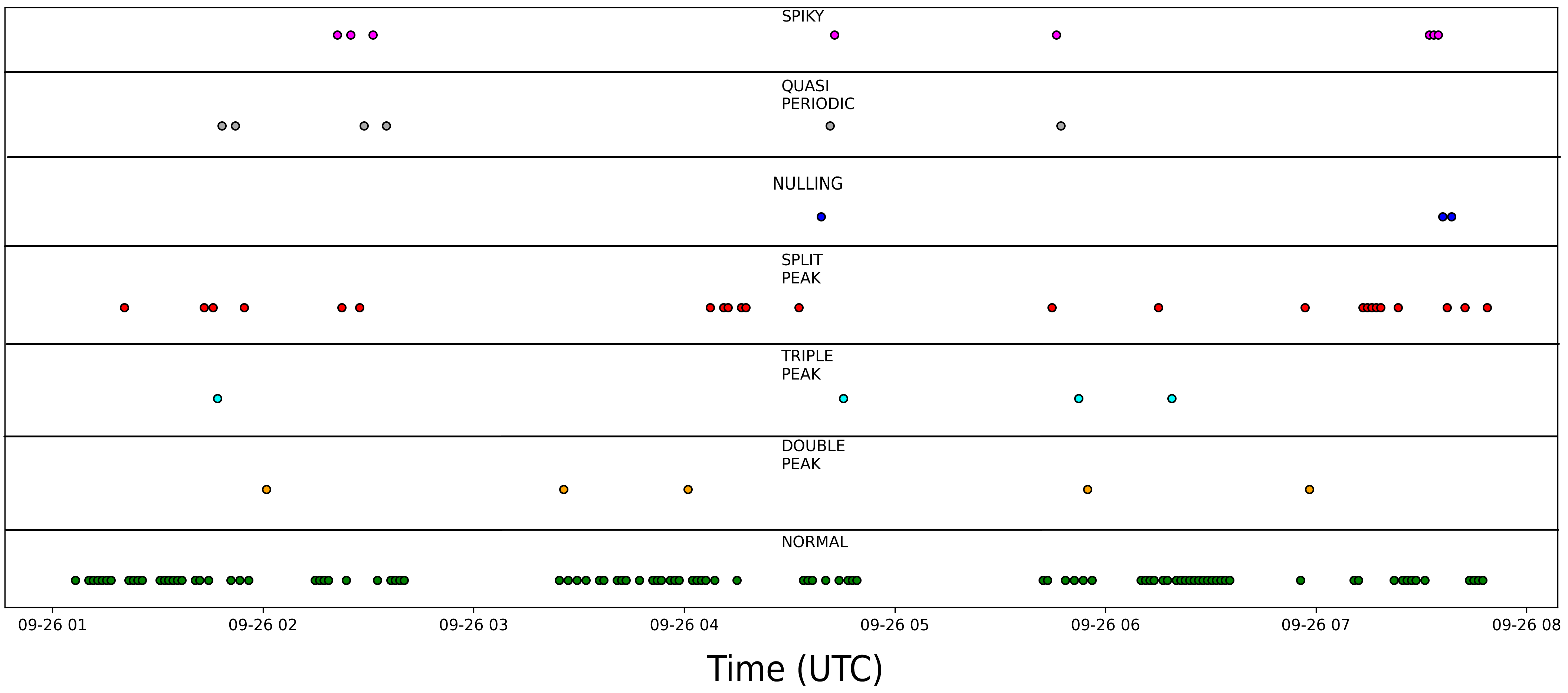}
    \caption[Times of PSR~J0901$-$4046 MeerKAT PTUSE detections divided by pulse shape type.]{Times of MeerKAT PTUSE detections of PSR~J0901$-$4046, categorised based on pulse shape type.}
    \label{fig:mtp13_mjd_shapes}
\end{figure*}

\subsubsection*{Pulse shape clustering}
Fig.\ \ref{fig:mtp13_mjd_shapes} shows the ToAs of all the detected single pulses categorised according to the pulse shape type. While \citetalias{2022caleb} found that pulse morphologies vary chaotically, the arrival times of certain morphological types in these data may suggest clustering in time. For example, eleven of the 24 \textit{split-peak} pulses occur in two separate ten-minute clusters (04:07--04:17 and 07:13--07:23), while three of the eight \textit{spiky} pulses occur in direct succession (07:32--07:34). To test for possible clustering of the pulse types, we used the method employed by \citet[][]{2018Oppermann} to identify temporal clustering in repeat pulses from FRB~20121102A. We fit a Weibull function,
\begin{equation}\label{eq:mtp13_weibull}
\mathcal{W}(\delta \mid k, r)=k \delta^{-1}[\delta r \Gamma(1+1 / k)]^{k} \mathrm{e}^{-[\delta r \Gamma(1+1 / k)]^{k}},
\end{equation}
where
\begin{equation}
\Gamma(x) = \int_{0}^{\infty} t^{x-1} \mathrm{e}^{-t} \mathrm{d} t 
\end{equation}
is the gamma function, to the distribution of wait times between subsequent pulses of the \textit{normal} type. If the pulse morphology corresponds to a Poissonian random process, then the distribution of wait times between subsequent pulses of the same type should follow an exponential function,
\begin{equation}\label{eq:mtp13_exp}
\mathcal{P}(\delta \mid r)=r \mathrm{e}^{-\delta r},
\end{equation}
where $r$ is the constant pulsation rate. The Weibull function in Eq.~\ref{eq:mtp13_weibull} equals an exponential function in the case that its shape parameter $k=1$. Hence, any deviation of the best-fit Weibull function's shape parameter from unity indicates clustering in the arrival times of the pulses.
\begin{figure}
    \centering
    \includegraphics[width=.48\textwidth]{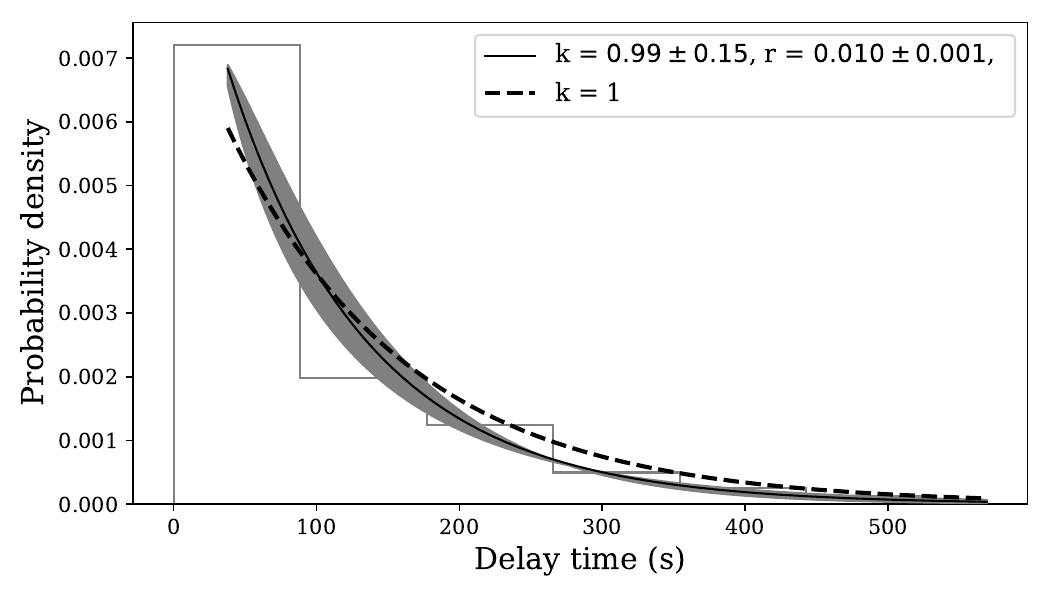}
    \caption[Probability density function of wait times between subsequent \textit{normal}-type PSR~J0901$-$4046 single pulses, with best-fit Weibull and exponential functions.]{Probability density function of wait times between subsequent PSR~J0901$-$4046 single pulses of the \textit{normal} type. The solid curve (with 1-$\sigma$ errors shaded) indicates the best-fit Weibull distribution with $k$ and the burst rate ($r$) as free parameters. The dashed curve shows the best-fit Weibull distribution assuming $k=1$, corresponding to an exponential function.}\label{fig:mtp13_weibull}
\end{figure}
The Weibull function fit to the wait times between subsequent \textit{normal} pulses from PSR~J0901$-$4046 are shown in Fig.~\ref{fig:mtp13_weibull}. The best-fit shape parameter $k$ is consistent with unity to within the 1-$\sigma$ errors. There is thus no evidence of temporal clustering of the \textit{normal} type pulses. The other morphological pulse types have too few examples for a similar analysis to be performed. It may be that the predominance of \textit{normal} type pulses in the data set makes any clustering difficult to discern, since the wait times will be dominated by those corresponding to one or two rotations of the pulsar. A similar analysis of the occurrence of a less prevalent pulse type may be more fruitful, but too few instances are available at present to be certain.

\subsection{Sub-pulse structure}
\begin{figure*}
    \centering
    \includegraphics[width=.99\textwidth]{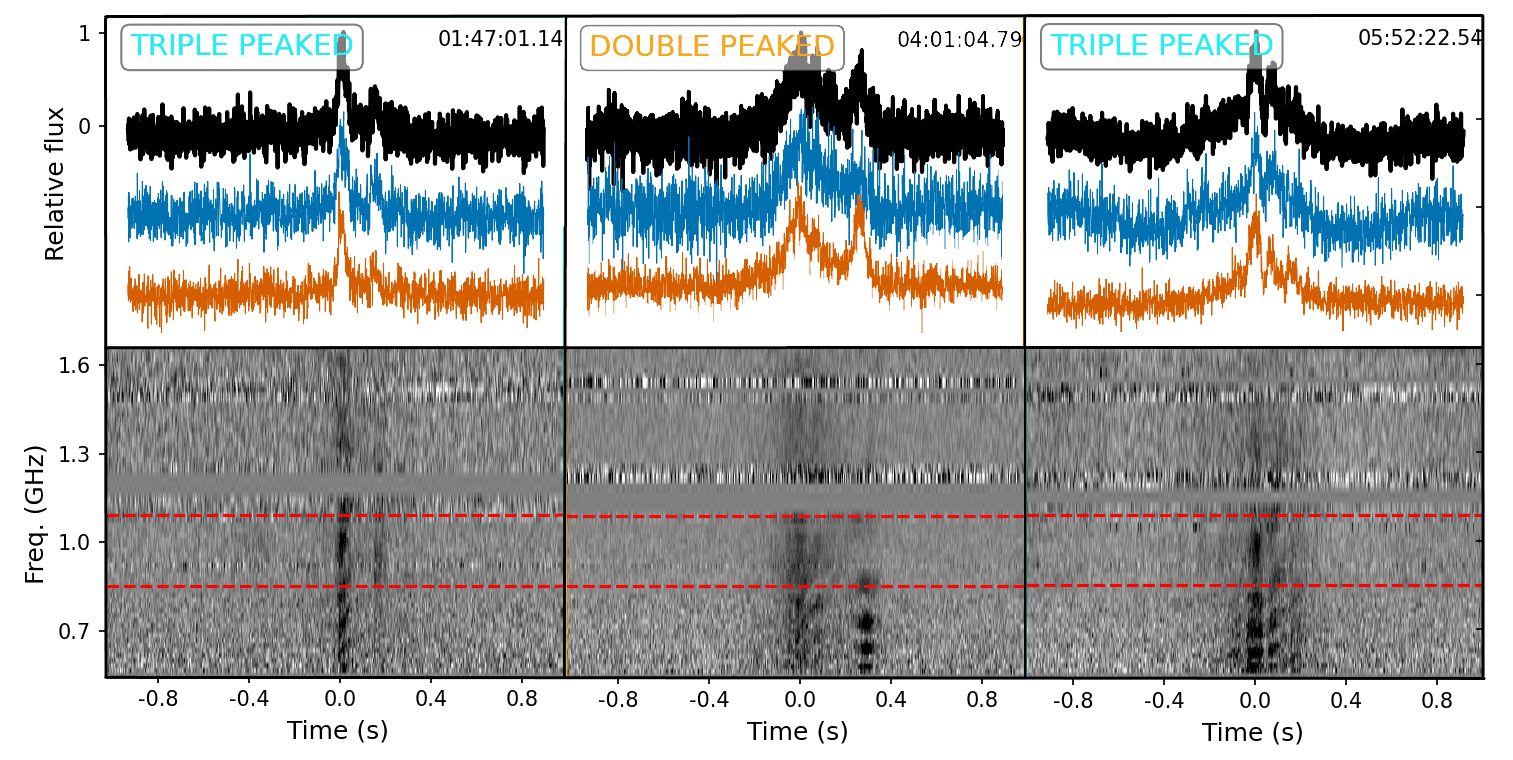}
    \caption{Profiles (top) and dynamic spectra (bottom) of three PSR~J0901$-$4046 pulses showing evidence of variable frequency-dependence of sub-pulses. Black profiles are averaged over the entire combined MeerKAT band, while the blue and orange profiles are averaged over the L-band and UHF-band only. The detection time (UTC) for each pulse is listed at the top of each panel.}\label{fig:sub-pulse_spectral_variance}
\end{figure*}

PSR~J0901$-$4046 single pulses of certain morphological types exhibit significant sub-pulse structure with varying spectro-temporal behaviour. For example, visual inspection of instances of the \textit{split peak}, \textit{double peak}, and \textit{triple peak} pulses suggests that their sub-pulse components may exhibit different spectral indices from the main pulse. The three pulses shown in Fig.\ \ref{fig:sub-pulse_spectral_variance}, two of which are \textit{triple-peak} and one \textit{double-peak}, are examples of pulses for which the sub-pulse components are visible over a notably restricted frequency band compared to the main pulse. In these cases, the sub-pulse components are clearly more visible in the L-band profile (blue) than the UHF-band profile or vice versa. On the other hand, the many instances of these same pulse types do not show any obvious deviation between the frequency dependence of the sub-pulse components. More detailed spectral fitting of these components is required to establish a definite trend.

Visual inspection of the single pulses also suggests that both the sub-pulse components and microstructure (i.e.~spikes, partial nulling, and quasi-periodic oscillations superimposed on the pulse), appear consistent in phase across the full frequency range available. This can be seen clearly in Fig.~\ref{fig:mtp13_shape_examples}, where the sub-pulse components and microstructure in the L band and UHF band are well aligned. Consistent with the absence of frequency-dependence found in $\S$\ref{subsec:pwidth}, we do not see clear evidence of sub-pulses drifting downward in frequency, which characterises some pulsars' and FRBs' emission \citep[see e.g.][]{2019hessels} in any of the single pulses in this data set. However, the average pulse has a secondary component that separates out at low frequencies (see Fig.~\ref{fig:mtp13_ptuse_sum}), which may indicate that this effect is weakly present. The pulse structure features also seem to have consistent widths across the full frequency band, although the data do not possess sufficient time resolution to confidently measure their widths.

In order to determine the time scales of the sub-pulses and microstructure present in some single pulses, we performed an auto-correlation function (ACF) analysis on their intensities \citep[see e.g.][]{1979cordes}. The time-dependent ACF is given by the cross-correlation of a given de-dispersed time series $f(t)$ with a copy of itself with some offset $\tau$, i.e.~
\begin{equation}
\mathrm{ACF}\left(\tau\right) = \int^{t}_0 f(t)f(t-\tau)\mathrm{dt}.
\end{equation}\label{eq:mtp13_acf}
The ACF will have a peak at $\tau=0$, corresponding to self-noise. A second peak at short time scales is indicative of the characteristic time scale of quasi-periodic microstructure.

\begin{figure*}
    \centering
    \includegraphics[width=.9\textwidth]{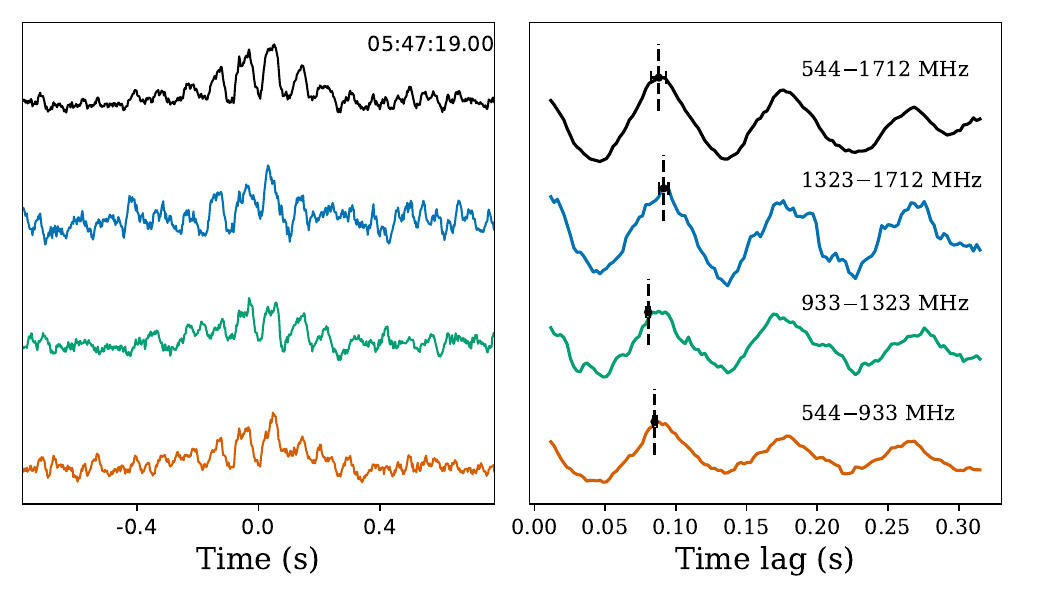}
    \caption[Auto-correlation function of a quasi-periodic PSR~J0901$-$4046 pulse.]{Pulse profiles (left) and ACFs (right) of a representative quasi-periodic PSR~J0901$-$4046 pulse across the full band and three sub-bands. The dashed vertical lines in the ACF plot indicate the peak of the ACF in each sub-band, and thus the quasi-period, with 1-$\sigma$ error bars.}\label{fig:mtp13_acf}
\end{figure*}

Fig.~\ref{fig:mtp13_acf} shows an example ACF for the \textit{quasi-periodic} pulse in Fig.~\ref{fig:mtp13_shape_examples}. The ACF is shown for the pulse profile averaged over the full combined band, as well as for three sub-bands\footnote{We found that three frequency subdivisions was the maximum number that still resulted in high confidence in the measured quasiperiod.}. The dashed vertical lines in the bottom panel indicate the peak of the ACF for each profile (disregarding the self-noise component), and thus the characteristic time scales of the quasi-periodic microstructure. We determined 1-$\sigma$ uncertainties on the ACF peak using bootstrap re-sampling with 10000 iterations. The ACF peaks were found to be at $\sim$90~ms for the full band (544--1712~MHz), $\sim$88~ms in the 544--933~MHz range, $\sim$83~ms in the 933--1323~MHz range, and $\sim$93~ms in the 1323--1717~MHz range. The time scale of the quasi-periodicity in this instance is thus highly stable across the combined band.

Using the ACF, we identified five pulses whose profiles exhibit quasi-periodic substructure. The characteristic time scales of the microstructure for these pulses are listed in Table~\ref{table:MTP13_quasiper_delays}. The time scales are given for both the full, combined band and three sub-bands. Division into more sub-bands results in signals in some sub bands that are too faint to perform the ACF analysis. The mean time scale over the full band is 72.86~ms, corresponding to roughly $(P/s)\times 10^{-3}$, with the means over the sub-bands all being within 2~ms of that value. It is evident that the characteristic time scales for all the \textit{quasi-periodic} pulses is roughly consistent across the frequency band.

We also applied the ACF analysis to the eight \textit{spiky} pulses in our data set, and for four of them found convincing quasi-periods, as listed in Table~\ref{table:MTP13_quasiper_delays}. These time scales were significantly shorter than those for the \textit{quasi-periodic} pulses, with a mean over the full band of 20.87~ms. This quasi-period also seems to remain approximately constant across the bandwidth.


\section{Discussion and conclusions}\label{sec:discussion}
In this paper, we have presented analytical results of a long term follow-up campaign of the ultra-long period magnetar PSR~J0901$-$4046 using four different telescopes. Pulses were observed from 544 to 4032~MHz, with the pulsar still strongly visible at the top of the Murriyang UWL band, but non-detection using both GMRT Band-3 and MWA may indicate a turnover in the spectral index below $\sim$500~MHz. Such low-frequency turnovers have been observed in a small subset of radio pulsars, usually $\sim$50--100~MHz \citep[][]{2018jankowski}, but turn-overs at $\geq$500~MHz have also been observed and attributed to free-free absorption in the interstellar medium \citep[e.g.][]{2017kijak}.

\begin{table*}
\centering
\caption[Characteristic time scales of microstructure in quasi-periodic PSR~J0901$-$4046 single pulses.]{Quasi-periodic microstructure characteristic time scales, measured across the full combined MeerKAT band as well as three sub-bands. Sub-bands 1, 2 and 3 represent 544--933~MHz, 933--1323~MHz and 1323--1712~MHz, respectively. The profiles of the \textit{spiky} pulses contain 0.4~seconds of data, while those of the \textit{quasi-periodic} pulses contain 1.0~second of data; the derived quasi-periods are indicated by blue vertical lines on the profiles.}\label{table:MTP13_quasiper_delays}

\begin{tabular}{lccrrrr}
\toprule
 & & & \multicolumn{4}{c}{ACF peak time (ms)}\\
\cmidrule(lr){4-7}
Pulse type & Profile & Arrival time & Full band & Sub-band 1 & Sub-band 2 & Sub-band 3\\
\midrule
Quasi-periodic & \raisebox{-\totalheight}{\includegraphics[width=0.2\textwidth, height=7mm]{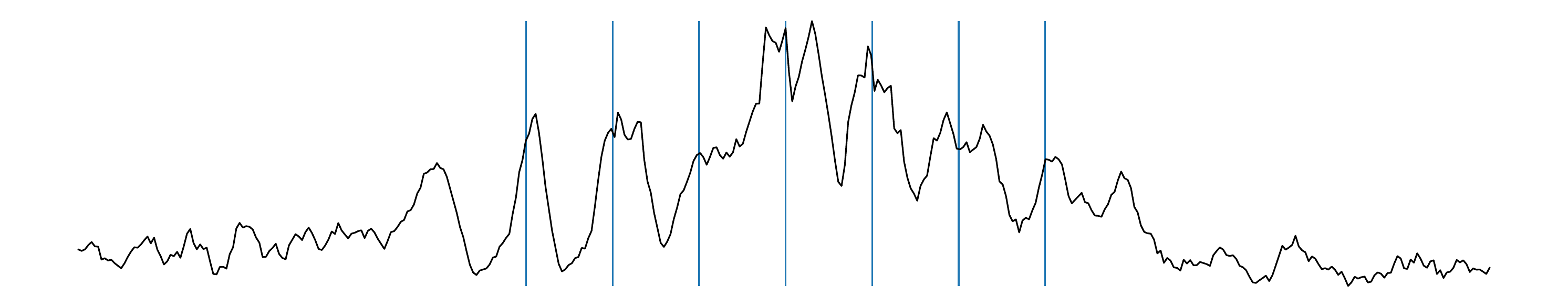}}  & 01:48:17.02 & 60.894(2) & 56.790(3) & 60.210(1) & 62.135(1)\\
Quasi-periodic & \raisebox{-\totalheight}{\includegraphics[width=0.2\textwidth, height=7mm]{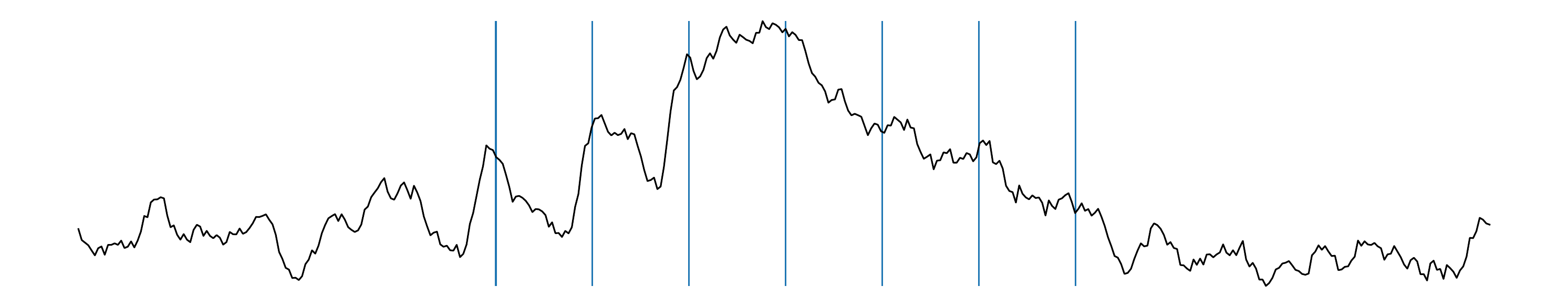}}  & 01:52:04.67 & 68.269(2) & 64.526(1) & 70.015(2) & 71.3254(6)\\
Quasi-periodic &  \raisebox{-\totalheight}{\includegraphics[width=0.2\textwidth, height=7mm]{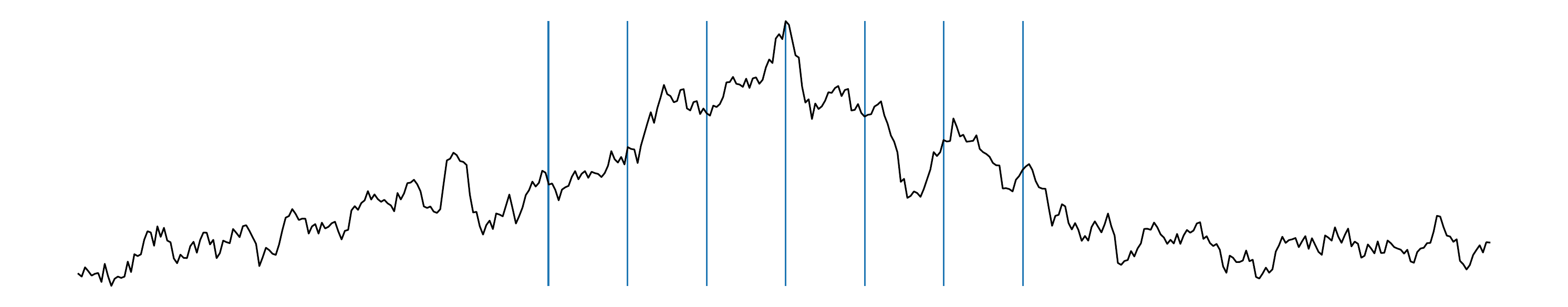}} & 02:35:04.91 & 55.6795(1) & 55.5741(1) & 56.978(4) & 56.510(3) \\
Quasi-periodic &  \raisebox{-\totalheight}{\includegraphics[width=0.2\textwidth, height=7mm]{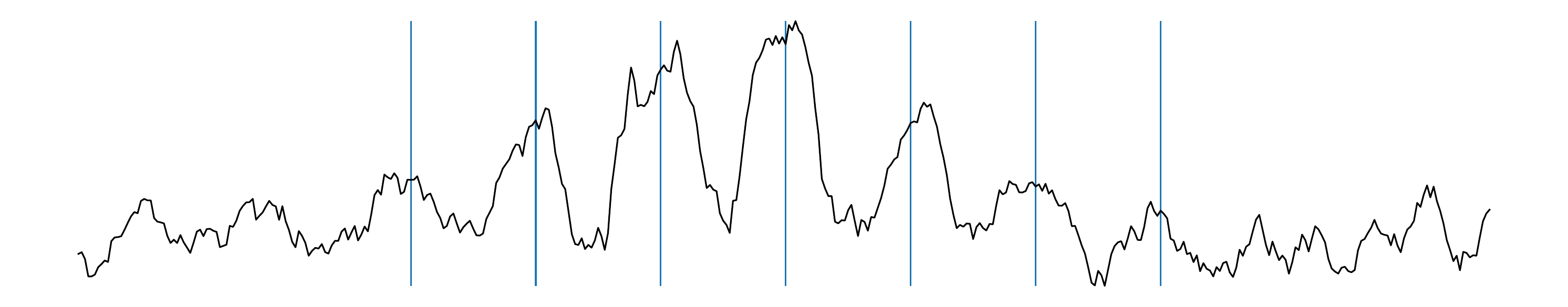}}  & 05:47:19.00 & 87.949(6) & 87.730(5) & 86.368(6) & 90.338(6)\\
Quasi-periodic &  \raisebox{-\totalheight}{\includegraphics[width=0.2\textwidth, height=7mm]{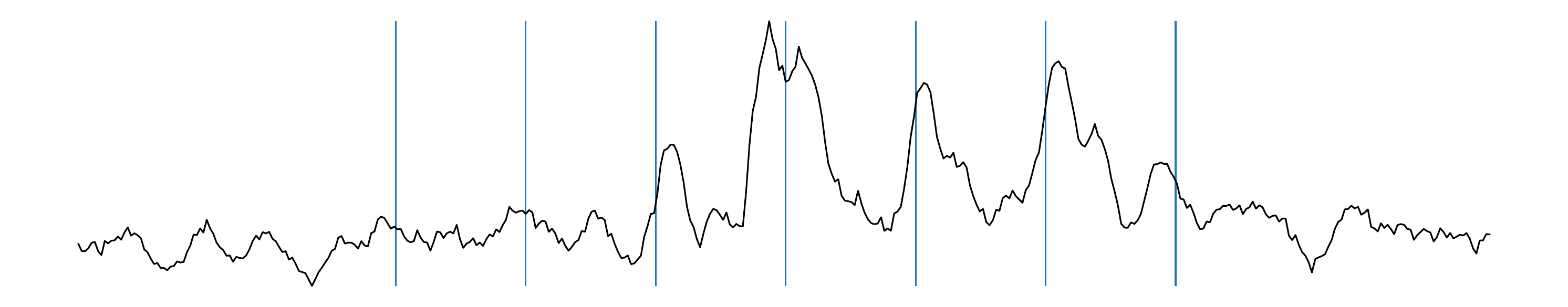}} & 06:18:56.09 & 91.510(4) & 95.598(3) & 90.0487(8) & 93.629(6)\\
\bottomrule
\textbf{Mean (Quasi-periodic)} & & & 72.86 &  72.04 & 72.72 & 74.79  \\
\bottomrule
Spiky & \raisebox{-\totalheight}{\includegraphics[width=0.2\textwidth, height=7mm]{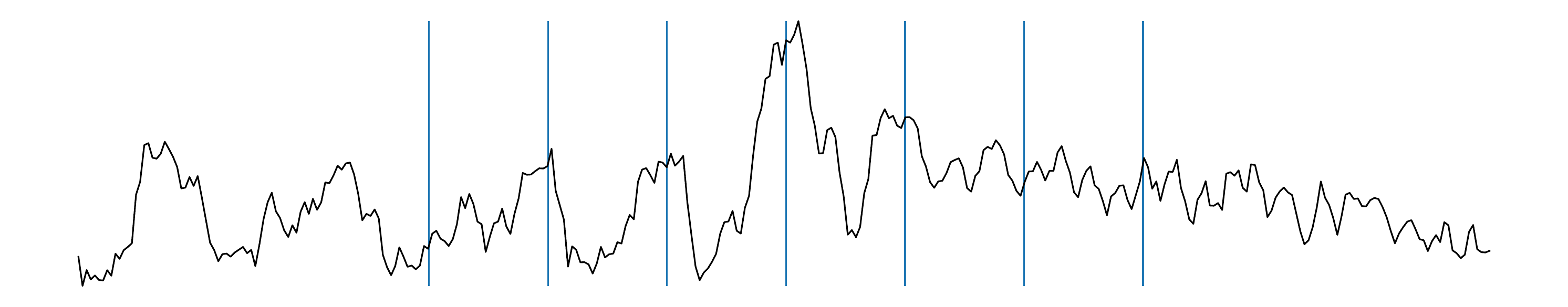}}  & 02:21:10.19 & 33.492(3) & 33.883(2) & 33.028(3) & 39.136(2)\\
Spiky & \raisebox{-\totalheight}{\includegraphics[width=0.2\textwidth, height=7mm]{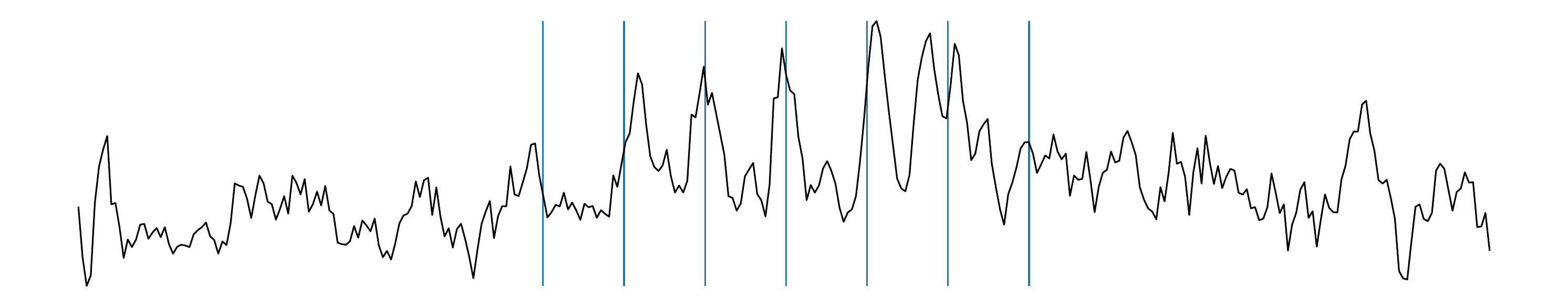}}  & 02:31:17.26 & 22.797(1) & 26.738(1) & 23.1115(3) & 20.384(1)\\
Spiky & \raisebox{-\totalheight}{\includegraphics[width=0.2\textwidth, height=7mm]{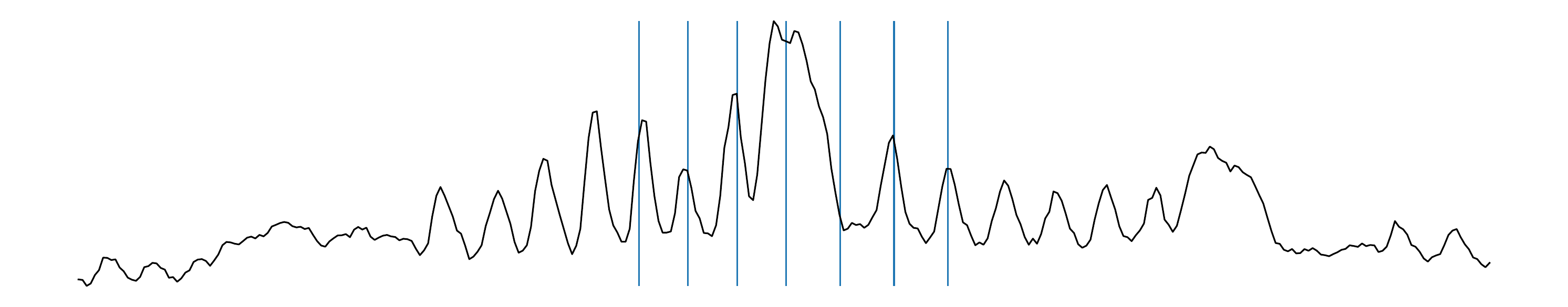}}  & 07:32:17.34 & 13.842(2) & 13.3393(6) & 14.3300(6) & 14.3323(6)\\
Spiky & \raisebox{-\totalheight}{\includegraphics[width=0.2\textwidth, height=7mm]{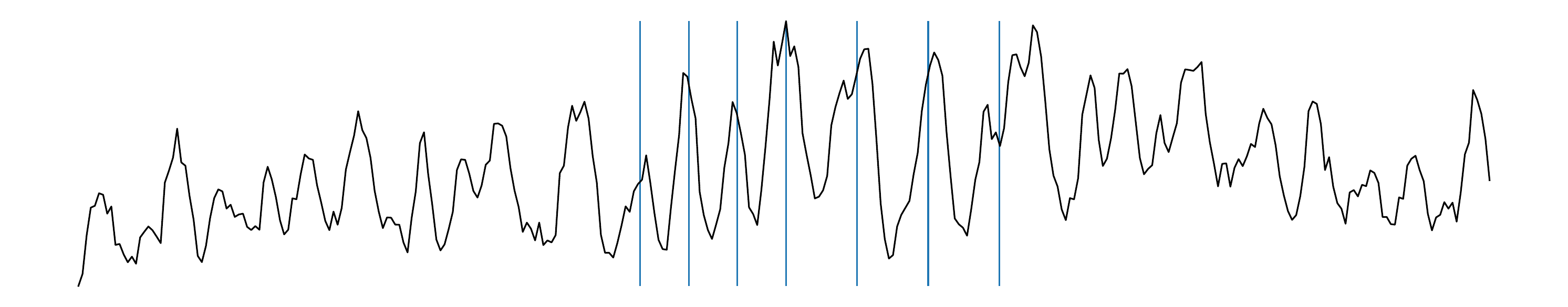}} & 07:33:33.22 & 13.346(1) & 13.711(1) & 13.114(1) & 13.5234(1)\\
\bottomrule
\textbf{Mean (Spiky)} & & & 20.87 &  21.92 & 20.90 & 21.84  \\
\bottomrule
\end{tabular}
\end{table*}

We have found that the timing solution has been highly stable since the discovery of the pulsar on 27 September 2020. No anomalous timing behaviour has been observed, and the RMS of the timing model ($\sim$8$\mu$s) is minimal compared to the pulse period. This suggests that, despite the presence of structural modulation, the pulse envelope is very stable from pulse to pulse. This stability over a long period of time is contrary to the typical behaviour of magnetars, which are characterised by frequent outbursts and associated timing glitches \citep[see][]{2017kaspi}, as well as drastic changes in the spin-down rate \citep[e\.g.][]{2019levin}. Likewise, the lack of observable variation in pulsed flux density is dissimilar to the drastic increases in flux that tend to accompany these anomalies in known radio magnetars. 

As outlined in $\S\ref{subsec:spectrum}$, we found that PSR~J0901$-$4046 possesses a significantly flatter average spectrum compared to regular radio pulsars over a wide range of frequencies, with the measured spectral index around $-1.0$, compared to the radio pulsar average of $\alpha=-1.6 \pm 0.03$ \citep[][]{2018jankowski}. However, the spectrum is still steeper than what has been observed in radio-loud magnetars, which have spectral indices ranging from $-0.5$ to $+0.9$ \citep[][]{2007camilo,2007camiloA,2008lazaridis,2010levin}.

Intriguingly, in $\S\ref{subsec:pwidth}$ we showed that the width of the averaged pulse of PSR~J0901$-$4046 remains steady over a wide frequency range, meaning that emission at various altitudes above the NS surface extends over roughly the same angular distance, which may be surprising given the expected size of the pulsar magnetosphere. One possible explanation is that, due to the very large light cylinder radius $R_{\mathrm{LC}}=cP/2\pi = 3.62\times10^{6}$~km, open magnetic field lines along which charged particles are accelerated possess a very low radius of curvature. Radiation beams emitted at various heights above the polar cap, corresponding to various emission frequencies, should therefore have very similar opening angles. Alternatively, if RFM-like behaviour is the result of a propagation effect such as birefringence in the magnetosphere, as has occasionally been suggested \citep[see e.g.][]{1996mckinnon,2022philippov}, this result may be used to probe the plasma properties in extreme magnetospheres. The absence of frequency-dependent behaviour in this pulsar is also notably in contrast to the phenomenon of frequency drift in the sub-bursts of repeating FRBs, which has often been explained by reference to RFM in a NS magnetosphere \citep[][]{2019hessels,2019wang}.

The steadiness of PSR~J0901$-$4046's pulse width is contrary to that of the next-slowest known radio pulsar, PSR~J0250+5854 ($P=23.5$~s), whose pulses widen appreciably at higher frequencies ($\mu=1.9\pm0.4$) \citep{2021agar}. However, in that pulsar's case, broadening is likely due to the frequency-dependence of secondary profile components. This is in agreement with the finding of \citet{2016pilia} in their study of 100 pulsars observed by LOFAR that profile broadening with higher frequency or non-monotonic behaviour was a result of the appearance/disappearance of profile components. A search of the literature did not provide measurements indicating RFM for any other radio pulsars with P$>5$~s, and so whether longer-period pulsars have less pronounced RFM is still uncertain.

We calculated PEDs for pulses detected in the MeerKAT L-band and UHF-band, which were both well-fit by log-normal distributions ($\S\ref{sec:penergy}$). We confirmed the finding in \citep[][]{2022caleb} that the pulse-to-pulse modulation is greater at lower frequencies. This is likely linked to the frequency-dependent behaviour of sub-pulse components that differs from that of the main pulse (see Fig.~\ref{fig:sub-pulse_spectral_variance}). Although we see instances of sub-pulse components being brighter in the L band compared to the UHF band, the disparate modulation index seems to indicate that the aggregate trend is of increased sub-pulse activity towards lower frequencies. Taking the typical interpretation that lower-frequency radio emission is emitted closer to the NS surface than higher-frequency emission, this indicates that the process responsible for these components is more active at lower altitudes. This behaviour is contrary to predictions that sub-pulse components should be largely consistent across large frequency ranges \citep[e.g.][]{1979cordes,1998lange}. Instances have been recorded of sub-pulse features widening towards lower frequencies, such as in the Vela pulsar \citep[][]{2002kramer}. \citetalias{2022caleb} used a continuous wavelet transform (CWT) analysis to show that the width of PSR~J0901$-$4046's sub-pulse components is stable over frequency. For the pulses in our sample, we also do not find any variation in the sub-pulse widths across a wide band. 

The high time-resolution average pulse profile and dynamic spectrum in Fig.~\ref{fig:mtp13_ptuse_sum} confirms the conclusion in \citetalias{2022caleb} that the average pulse shape is different at lower frequencies than at higher frequencies, with a distinct secondary pulse component visible in the UHF band profile but absent from the L band profile. 
There is also evidence in many of the observed single pulses that sub-pulse features apparently have different frequency dependencies than the main pulse. It may thus be that these sub-features originate from elsewhere in the radio emission region of the NS.

Interestingly, the pulse morphology analysis in $\S\ref{subsec:pmorph}$ has resulted in different proportions of each of the seven pulse types compared to that reported in \citep[][]{2022caleb}. Notably, the \textit{normal} type morphology dominates the single pulses that we have analysed.

ACF analysis of the microstructure of pulses exhibiting quasi-periodicity has shown that these features appear at roughly consistent points in the pulse phase across frequencies. A larger sample of such pulses is required to ascertain whether this is true for all single pulses and for all types of microstructure. The ACF analysis of \citet[][]{2022caleb} found quasi-periods in 25 pulses averaging 75.82~ms, remarkably close to the $\sim P\times10^{-3}$ scaling proposed as common NS behaviour by \citet{2024kramer}. It should be noted, however, that the quasi-period varied widely between single pulses, from about 10~ms to 340~ms. The quasi-periods we measure for five pulses in this data set are all within that range, with an average over the full frequency range of 61.14~ms. \citet[][]{2024kramer} interprets the scatter of the quasi-period as a result of fluctuations about a preferred value, and notes that the relationship is robust despite large uncertainties on quasi-period measurements for NSs with rotation periods spanning six orders of magnitude.

\section*{Acknowledgements}

The MeerKAT telescope is operated by the South African Radio Astronomy Observatory, which is a facility of the National Research Foundation, an agency of the Department of Science and Innovation (DSI). The FBFUSE and APSUSE computer clusters used  for data acquisition, storage and analysis were funded and installed by the Max-Planck-Institut für Radioastronomie and the Max-PlanckGesellschaft. Murriyang, CSIRO’s Parkes radio telescope, is part of the \href{https://ror.org/05qajvd42}{Australia Telescope National Facility} which is funded by the Australian Government for operation as a National Facility managed by CSIRO. We acknowledge the Wiradjuri people as the Traditional Owners of the Observatory site. GMRT is run by the National Centre for Radio Astrophysics of the Tata Institute of Fundamental Research. We gratefully acknowledge the Department of Atomic Energy, Government of India, for its assistance under project No. 12-R\&D-TFR-5.02-0700. This scientific work uses data obtained from Inyarrimanha Ilgari Bundara, the CSIRO Murchison Radio-astronomy Observatory. We acknowledge the Wajarri Yamaji People as the Traditional Owners and Native Title Holders of the Observatory site. Support for the operation of the MWA is provided by the Australian Government (NCRIS), under a contract to Curtin University administered by Astronomy Australia Limited. We acknowledge the Pawsey Supercomputing Research Centre which is supported by the Western Australian and Australian Governments. The MeerTRAP collaboration would like to thank the MeerKAT Large Survey Project teams for allowing MeerTRAP to observe commensally. The MeerTRAP collaboration acknowledges funding from the European Research Council (ERC) under the European Union’s Horizon 2020 research and innovation programme (grant agreement No 694745). B.W.S. acknowledges funding from a UK Science and Technology Facilities Council (STFC) consolidated grant. M.C. acknowledges support of an Australian Research Council Discovery Early Career Research Award (project number DE220100819) funded by the Australian Government. Parts of this research were conducted by the Australian Research Council Centre of Excellence for Gravitational Wave Discovery (OzGrav), project number CE170100004. K.M.R. acknowledges support from the Vici research program ’ARGO’ with project number 639.043.815, financed by the Dutch Research Council (NWO). L.N.D would like to acknowledge the Gadigal People of the Eora nation, the traditional owners of the land on which she works and lives.
\section*{Data Availability}

The data will be made available to others upon reasonable request to
the authors.



\bibliographystyle{mnras}
\bibliography{main} 

\begin{thebibliography}{}
\makeatletter
\relax
\def\mn@urlcharsother{\let\do\@makeother \do\$\do\&\do\#\do\^\do\_\do\%\do\~}
\def\mn@doi{\begingroup\mn@urlcharsother \@ifnextchar [ {\mn@doi@} {\mn@doi@[]}}
\def\mn@doi@[#1]#2{\def\@tempa{#1}\ifx\@tempa\@empty \href {http://dx.doi.org/#2} {doi:#2}\else \href {http://dx.doi.org/#2} {#1}\fi \endgroup}
\def\mn@eprint#1#2{\mn@eprint@#1:#2::\@nil}
\def\mn@eprint@arXiv#1{\href {http://arxiv.org/abs/#1} {{\tt arXiv:#1}}}
\def\mn@eprint@dblp#1{\href {http://dblp.uni-trier.de/rec/bibtex/#1.xml} {dblp:#1}}
\def\mn@eprint@#1:#2:#3:#4\@nil{\def\@tempa {#1}\def\@tempb {#2}\def\@tempc {#3}\ifx \@tempc \@empty \let \@tempc \@tempb \let \@tempb \@tempa \fi \ifx \@tempb \@empty \def\@tempb {arXiv}\fi \@ifundefined {mn@eprint@\@tempb}{\@tempb:\@tempc}{\expandafter \expandafter \csname mn@eprint@\@tempb\endcsname \expandafter{\@tempc}}}

\bibitem[\protect\citeauthoryear{{Agar} et~al.,}{{Agar} et~al.}{2021}]{2021agar}
{Agar} C.~H.,  et~al., 2021, \mn@doi [\mnras] {10.1093/mnras/stab2496}, \href {https://ui.adsabs.harvard.edu/abs/2021MNRAS.508.1102A} {508, 1102}

\bibitem[\protect\citeauthoryear{{Bailes} et~al.,}{{Bailes} et~al.}{2020}]{2020bailes}
{Bailes} M.,  et~al., 2020, \mn@doi [\pasa] {10.1017/pasa.2020.19}, \href {https://ui.adsabs.harvard.edu/abs/2020PASA...37...28B} {37, e028}

\bibitem[\protect\citeauthoryear{{Baring} \& {Harding}}{{Baring} \& {Harding}}{1998}]{1998baring}
{Baring} M.~G.,  {Harding} A.~K.,  1998, \mn@doi [\apjl] {10.1086/311679}, \href {https://ui.adsabs.harvard.edu/abs/1998ApJ...507L..55B} {507, L55}

\bibitem[\protect\citeauthoryear{{Becker}, {Kramer}  \& {Sesana}}{{Becker} et~al.}{2018}]{2018becker}
{Becker} W.,  {Kramer} M.,   {Sesana} A.,  2018, \mn@doi [\ssr] {10.1007/s11214-017-0459-0}, \href {https://ui.adsabs.harvard.edu/abs/2018SSRv..214...30B} {214, 30}

\bibitem[\protect\citeauthoryear{{Beniamini}, {Wadiasingh}  \& {Metzger}}{{Beniamini} et~al.}{2020}]{2020beniamini}
{Beniamini} P.,  {Wadiasingh} Z.,   {Metzger} B.~D.,  2020, \mn@doi [\mnras] {10.1093/mnras/staa1783}, \href {https://ui.adsabs.harvard.edu/abs/2020MNRAS.496.3390B} {496, 3390}

\bibitem[\protect\citeauthoryear{Bezuidenhout}{Bezuidenhout}{2022}]{2022bezuidenhoutThesis}
Bezuidenhout M.~C.,  2022, PhD thesis, University of Manchester

\bibitem[\protect\citeauthoryear{{Bochenek}, {Ravi}, {Belov}, {Hallinan}, {Kocz}, {Kulkarni}  \& {McKenna}}{{Bochenek} et~al.}{2020}]{2020bochenek}
{Bochenek} C.~D.,  {Ravi} V.,  {Belov} K.~V.,  {Hallinan} G.,  {Kocz} J.,  {Kulkarni} S.~R.,   {McKenna} D.~L.,  2020, \mn@doi [\nat] {10.1038/s41586-020-2872-x}, \href {https://ui.adsabs.harvard.edu/abs/2020Natur.587...59B} {587, 59}

\bibitem[\protect\citeauthoryear{{CHIME/FRB Collaboration} et~al.,}{{CHIME/FRB Collaboration} et~al.}{2020}]{2020chime}
{CHIME/FRB Collaboration} et~al., 2020, \mn@doi [\nat] {10.1038/s41586-020-2863-y}, \href {https://ui.adsabs.harvard.edu/abs/2020Natur.587...54C} {587, 54}

\bibitem[\protect\citeauthoryear{{CHIME/FRB Collaboration} et~al.,}{{CHIME/FRB Collaboration} et~al.}{2022}]{2021chime}
{CHIME/FRB Collaboration} et~al., 2022, \mn@doi [\nat] {10.1038/s41586-022-04841-8}, \href {https://ui.adsabs.harvard.edu/abs/2022Natur.607..256C} {607, 256}

\bibitem[\protect\citeauthoryear{{Caleb} et~al.,}{{Caleb} et~al.}{2022a}]{2022caleb}
{Caleb} M.,  et~al., 2022a, \mn@doi [Nature Astronomy] {10.1038/s41550-022-01688-x}, \href {https://ui.adsabs.harvard.edu/abs/2022NatAs...6..828C} {6, 828}

\bibitem[\protect\citeauthoryear{{Caleb} et~al.,}{{Caleb} et~al.}{2022b}]{2022caleb2}
{Caleb} M.,  et~al., 2022b, \mn@doi [\mnras] {10.1093/mnras/stab3223}, \href {https://ui.adsabs.harvard.edu/abs/2022MNRAS.510.1996C} {510, 1996}

\bibitem[\protect\citeauthoryear{{Caleb} et~al.,}{{Caleb} et~al.}{2024}]{2024caleb}
{Caleb} M.,  et~al., 2024, \mn@doi [Nature Astronomy] {10.1038/s41550-024-02277-w}, \href {https://ui.adsabs.harvard.edu/abs/2024NatAs...8.1159C} {8, 1159}

\bibitem[\protect\citeauthoryear{{Camilo}, {Ransom}, {Halpern}, {Reynolds}, {Helfand}, {Zimmerman}  \& {Sarkissian}}{{Camilo} et~al.}{2006}]{2006camilo}
{Camilo} F.,  {Ransom} S.~M.,  {Halpern} J.~P.,  {Reynolds} J.,  {Helfand} D.~J.,  {Zimmerman} N.,   {Sarkissian} J.,  2006, \mn@doi [\nat] {10.1038/nature04986}, \href {https://ui.adsabs.harvard.edu/abs/2006Natur.442..892C} {442, 892}

\bibitem[\protect\citeauthoryear{{Camilo}, {Ransom}, {Halpern}  \& {Reynolds}}{{Camilo} et~al.}{2007a}]{2007camilo}
{Camilo} F.,  {Ransom} S.~M.,  {Halpern} J.~P.,   {Reynolds} J.,  2007a, \mn@doi [\apjl] {10.1086/521826}, \href {https://ui.adsabs.harvard.edu/abs/2007ApJ...666L..93C} {666, L93}

\bibitem[\protect\citeauthoryear{{Camilo} et~al.,}{{Camilo} et~al.}{2007b}]{2007camiloA}
{Camilo} F.,  et~al., 2007b, \mn@doi [\apj] {10.1086/521548}, \href {https://ui.adsabs.harvard.edu/abs/2007ApJ...669..561C} {669, 561}

\bibitem[\protect\citeauthoryear{{Camilo} et~al.,}{{Camilo} et~al.}{2018}]{2018camilo}
{Camilo} F.,  et~al., 2018, \mn@doi [\apj] {10.3847/1538-4357/aab35a}, \href {https://ui.adsabs.harvard.edu/abs/2018ApJ...856..180C} {856, 180}

\bibitem[\protect\citeauthoryear{{Chen} \& {Wang}}{{Chen} \& {Wang}}{2014}]{2014chen}
{Chen} J.~L.,  {Wang} H.~G.,  2014, \mn@doi [\apjs] {10.1088/0067-0049/215/1/11}, \href {https://ui.adsabs.harvard.edu/abs/2014ApJS..215...11C} {215, 11}

\bibitem[\protect\citeauthoryear{{Chen}, {Barr}, {Karuppusamy}, {Kramer}  \& {Stappers}}{{Chen} et~al.}{2021}]{2021chen}
{Chen} W.,  {Barr} E.,  {Karuppusamy} R.,  {Kramer} M.,   {Stappers} B.,  2021, \mn@doi [Journal of Astronomical Instrumentation] {10.1142/S2251171721500136}, \href {https://ui.adsabs.harvard.edu/abs/2021JAI....1050013C} {10, 2150013}

\bibitem[\protect\citeauthoryear{{Cordes}}{{Cordes}}{1978}]{1978cordes}
{Cordes} J.~M.,  1978, \mn@doi [\apj] {10.1086/156218}, \href {https://ui.adsabs.harvard.edu/abs/1978ApJ...222.1006C} {222, 1006}

\bibitem[\protect\citeauthoryear{{Cordes}}{{Cordes}}{1979}]{1979cordes}
{Cordes} J.~M.,  1979, \mn@doi [Australian Journal of Physics] {10.1071/PH790009}, \href {https://ui.adsabs.harvard.edu/abs/1979AuJPh..32....9C} {32, 9}

\bibitem[\protect\citeauthoryear{{Cordes} \& {Chatterjee}}{{Cordes} \& {Chatterjee}}{2019}]{2019cordes}
{Cordes} J.~M.,  {Chatterjee} S.,  2019, \mn@doi [\araa] {10.1146/annurev-astro-091918-104501}, \href {https://ui.adsabs.harvard.edu/abs/2019ARA&A..57..417C} {57, 417}

\bibitem[\protect\citeauthoryear{{Cordes} \& {Lazio}}{{Cordes} \& {Lazio}}{2002}]{ne2001}
{Cordes} J.~M.,  {Lazio} T.~J.~W.,  2002, \mn@doi [arXiv e-prints] {10.48550/arXiv.astro-ph/0207156}, \href {https://ui.adsabs.harvard.edu/abs/2002astro.ph..7156C} {pp astro--ph/0207156}

\bibitem[\protect\citeauthoryear{{De}, {Gupta}  \& {Sharma}}{{De} et~al.}{2016}]{2016de}
{De} K.,  {Gupta} Y.,   {Sharma} P.,  2016, \mn@doi [\apjl] {10.3847/2041-8213/833/1/L10}, \href {https://ui.adsabs.harvard.edu/abs/2016ApJ...833L..10D} {833, L10}

\bibitem[\protect\citeauthoryear{{Dong} et~al.,}{{Dong} et~al.}{2024}]{2024dong}
{Dong} F.~A.,  et~al., 2024, \mn@doi [arXiv e-prints] {10.48550/arXiv.2407.07480}, \href {https://ui.adsabs.harvard.edu/abs/2024arXiv240707480D} {p. arXiv:2407.07480}

\bibitem[\protect\citeauthoryear{{Esposito} et~al.,}{{Esposito} et~al.}{2020}]{2020esposito}
{Esposito} P.,  et~al., 2020, \mn@doi [\apjl] {10.3847/2041-8213/ab9742}, \href {https://ui.adsabs.harvard.edu/abs/2020ApJ...896L..30E} {896, L30}

\bibitem[\protect\citeauthoryear{{Fender} et~al.,}{{Fender} et~al.}{2016}]{2016fender}
{Fender} R.,  et~al., 2016, in MeerKAT Science: On the Pathway to the SKA. p.~13 (\mn@eprint {arXiv} {1711.04132}), \mn@doi{10.22323/1.277.0013}

\bibitem[\protect\citeauthoryear{{Gupta} et~al.,}{{Gupta} et~al.}{2017}]{2017gupta}
{Gupta} Y.,  et~al., 2017, \mn@doi [Current Science] {10.18520/cs/v113/i04/707-714}, \href {https://ui.adsabs.harvard.edu/abs/2017CSci..113..707G} {113, 707}

\bibitem[\protect\citeauthoryear{{Hessels} et~al.,}{{Hessels} et~al.}{2019}]{2019hessels}
{Hessels} J.~W.~T.,  et~al., 2019, \mn@doi [\apjl] {10.3847/2041-8213/ab13ae}, \href {https://ui.adsabs.harvard.edu/abs/2019ApJ...876L..23H} {876, L23}

\bibitem[\protect\citeauthoryear{{Heyl} \& {Hernquist}}{{Heyl} \& {Hernquist}}{2005}]{2005heyl}
{Heyl} J.~S.,  {Hernquist} L.,  2005, \mn@doi [\mnras] {10.1111/j.1365-2966.2005.09338.x}, \href {https://ui.adsabs.harvard.edu/abs/2005MNRAS.362..777H} {362, 777}

\bibitem[\protect\citeauthoryear{{Hobbs}, {Edwards}  \& {Manchester}}{{Hobbs} et~al.}{2006}]{2006hobbs}
{Hobbs} G.~B.,  {Edwards} R.~T.,   {Manchester} R.~N.,  2006, \mn@doi [\mnras] {10.1111/j.1365-2966.2006.10302.x}, \href {https://ui.adsabs.harvard.edu/abs/2006MNRAS.369..655H} {369, 655}

\bibitem[\protect\citeauthoryear{{Hobbs} et~al.,}{{Hobbs} et~al.}{2020}]{2020hobbs}
{Hobbs} G.,  et~al., 2020, \mn@doi [\pasa] {10.1017/pasa.2020.2}, \href {https://ui.adsabs.harvard.edu/abs/2020PASA...37...12H} {37, e012}

\bibitem[\protect\citeauthoryear{{Hotan}, {van Straten}  \& {Manchester}}{{Hotan} et~al.}{2004}]{2004Hotan}
{Hotan} A.~W.,  {van Straten} W.,   {Manchester} R.~N.,  2004, \mn@doi [\pasa] {10.1071/AS04022}, \href {https://ui.adsabs.harvard.edu/abs/2004PASA...21..302H} {21, 302}

\bibitem[\protect\citeauthoryear{{Hu} et~al.,}{{Hu} et~al.}{2020}]{2020hu}
{Hu} C.-P.,  et~al., 2020, \mn@doi [\apj] {10.3847/1538-4357/abb3c9}, \href {https://ui.adsabs.harvard.edu/abs/2020ApJ...902....1H} {902, 1}

\bibitem[\protect\citeauthoryear{{Hurley-Walker} et~al.,}{{Hurley-Walker} et~al.}{2022}]{2022hurleywalker}
{Hurley-Walker} N.,  et~al., 2022, \mn@doi [\nat] {10.1038/s41586-021-04272-x}, \href {https://ui.adsabs.harvard.edu/abs/2022Natur.601..526H} {601, 526}

\bibitem[\protect\citeauthoryear{{Hurley-Walker} et~al.,}{{Hurley-Walker} et~al.}{2023}]{2023hurleywalker}
{Hurley-Walker} N.,  et~al., 2023, \mn@doi [\nat] {10.1038/s41586-023-06202-5}, \href {https://ui.adsabs.harvard.edu/abs/2023Natur.619..487H} {619, 487}

\bibitem[\protect\citeauthoryear{{Hurley-Walker} et~al.,}{{Hurley-Walker} et~al.}{2024}]{2024hurleywalker}
{Hurley-Walker} N.,  et~al., 2024, \mn@doi [\apjl] {10.3847/2041-8213/ad890e}, \href {https://ui.adsabs.harvard.edu/abs/2024ApJ...976L..21H} {976, L21}

\bibitem[\protect\citeauthoryear{{Jankowski}, {van Straten}, {Keane}, {Bailes}, {Barr}, {Johnston}  \& {Kerr}}{{Jankowski} et~al.}{2018}]{2018jankowski}
{Jankowski} F.,  {van Straten} W.,  {Keane} E.~F.,  {Bailes} M.,  {Barr} E.~D.,  {Johnston} S.,   {Kerr} M.,  2018, \mn@doi [\mnras] {10.1093/mnras/stx2476}, \href {https://ui.adsabs.harvard.edu/abs/2018MNRAS.473.4436J} {473, 4436}

\bibitem[\protect\citeauthoryear{{Jankowski} et~al.,}{{Jankowski} et~al.}{2022}]{2020jankowski}
{Jankowski} F.,  et~al., 2022, in {Ruiz} J.~E.,  {Pierfedereci} F.,   {Teuben} P.,  eds,  Astronomical Society of the Pacific Conference Series Vol. 532, Astronomical Data Analysis Software and Systems XXX. p.~273 (\mn@eprint {arXiv} {2012.05173}), \mn@doi{10.48550/arXiv.2012.05173}

\bibitem[\protect\citeauthoryear{{Johnston}, {Karastergiou}, {Mitra}  \& {Gupta}}{{Johnston} et~al.}{2008}]{2008johnston}
{Johnston} S.,  {Karastergiou} A.,  {Mitra} D.,   {Gupta} Y.,  2008, \mn@doi [\mnras] {10.1111/j.1365-2966.2008.13379.x}, \href {https://ui.adsabs.harvard.edu/abs/2008MNRAS.388..261J} {388, 261}

\bibitem[\protect\citeauthoryear{{Jonas} \& {MeerKAT Team}}{{Jonas} \& {MeerKAT Team}}{2016}]{2016jonas}
{Jonas} J.,  {MeerKAT Team} 2016, in MeerKAT Science: On the Pathway to the SKA. p.~1, \mn@doi{10.22323/1.277.0001}

\bibitem[\protect\citeauthoryear{{Kaspi} \& {Beloborodov}}{{Kaspi} \& {Beloborodov}}{2017}]{2017kaspi}
{Kaspi} V.~M.,  {Beloborodov} A.~M.,  2017, \mn@doi [\araa] {10.1146/annurev-astro-081915-023329}, \href {https://ui.adsabs.harvard.edu/abs/2017ARA&A..55..261K} {55, 261}

\bibitem[\protect\citeauthoryear{{Kijak}, {Kramer}, {Wielebinski}  \& {Jessner}}{{Kijak} et~al.}{1998}]{1998kijak}
{Kijak} J.,  {Kramer} M.,  {Wielebinski} R.,   {Jessner} A.,  1998, \mn@doi [\aaps] {10.1051/aas:1998340}, \href {https://ui.adsabs.harvard.edu/abs/1998A&AS..127..153K} {127, 153}

\bibitem[\protect\citeauthoryear{{Kijak}, {Basu}, {Lewandowski}, {Ro{\.z}ko}  \& {Dembska}}{{Kijak} et~al.}{2017}]{2017kijak}
{Kijak} J.,  {Basu} R.,  {Lewandowski} W.,  {Ro{\.z}ko} K.,   {Dembska} M.,  2017, \mn@doi [\apj] {10.3847/1538-4357/aa6ff2}, \href {https://ui.adsabs.harvard.edu/abs/2017ApJ...840..108K} {840, 108}

\bibitem[\protect\citeauthoryear{{Kramer}, {Johnston}  \& {van Straten}}{{Kramer} et~al.}{2002}]{2002kramer}
{Kramer} M.,  {Johnston} S.,   {van Straten} W.,  2002, \mn@doi [\mnras] {10.1046/j.1365-8711.2002.05478.x}, \href {https://ui.adsabs.harvard.edu/abs/2002MNRAS.334..523K} {334, 523}

\bibitem[\protect\citeauthoryear{{Kramer}, {Liu}, {Desvignes}, {Karuppusamy}  \& {Stappers}}{{Kramer} et~al.}{2024}]{2024kramer}
{Kramer} M.,  {Liu} K.,  {Desvignes} G.,  {Karuppusamy} R.,   {Stappers} B.~W.,  2024, \mn@doi [Nature Astronomy] {10.1038/s41550-023-02125-3}, \href {https://ui.adsabs.harvard.edu/abs/2024NatAs...8..230K} {8, 230}

\bibitem[\protect\citeauthoryear{{Lange}, {Kramer}, {Wielebinski}  \& {Jessner}}{{Lange} et~al.}{1998}]{1998lange}
{Lange} C.,  {Kramer} M.,  {Wielebinski} R.,   {Jessner} A.,  1998, \aap, \href {https://ui.adsabs.harvard.edu/abs/1998A&A...332..111L} {332, 111}

\bibitem[\protect\citeauthoryear{{Lazaridis}, {Jessner}, {Kramer}, {Stappers}, {Lyne}, {Jordan}, {Serylak}  \& {Zensus}}{{Lazaridis} et~al.}{2008}]{2008lazaridis}
{Lazaridis} K.,  {Jessner} A.,  {Kramer} M.,  {Stappers} B.~W.,  {Lyne} A.~G.,  {Jordan} C.~A.,  {Serylak} M.,   {Zensus} J.~A.,  2008, \mn@doi [\mnras] {10.1111/j.1365-2966.2008.13794.x}, \href {https://ui.adsabs.harvard.edu/abs/2008MNRAS.390..839L} {390, 839}

\bibitem[\protect\citeauthoryear{{Lee} et~al.,}{{Lee} et~al.}{2025}]{2025lee}
{Lee} Y.~W.~J.,  et~al., 2025, \mn@doi [arXiv e-prints] {10.48550/arXiv.2501.09133}, \href {https://ui.adsabs.harvard.edu/abs/2025arXiv250109133L} {p. arXiv:2501.09133}

\bibitem[\protect\citeauthoryear{{Levin} et~al.,}{{Levin} et~al.}{2010}]{2010levin}
{Levin} L.,  et~al., 2010, \mn@doi [\apjl] {10.1088/2041-8205/721/1/L33}, \href {https://ui.adsabs.harvard.edu/abs/2010ApJ...721L..33L} {721, L33}

\bibitem[\protect\citeauthoryear{{Levin} et~al.,}{{Levin} et~al.}{2019}]{2019levin}
{Levin} L.,  et~al., 2019, \mn@doi [\mnras] {10.1093/mnras/stz2074}, \href {https://ui.adsabs.harvard.edu/abs/2019MNRAS.488.5251L} {488, 5251}

\bibitem[\protect\citeauthoryear{{Lorimer}}{{Lorimer}}{2011}]{2011lorimer}
{Lorimer} D.~R.,  2011, {SIGPROC: Pulsar Signal Processing Programs}, Astrophysics Source Code Library, record ascl:1107.016

\bibitem[\protect\citeauthoryear{{Lorimer} \& {Kramer}}{{Lorimer} \& {Kramer}}{2012}]{2012lorimer}
{Lorimer} D.~R.,  {Kramer} M.,  2012, {Handbook of Pulsar Astronomy}.
{Cambridge University Press}

\bibitem[\protect\citeauthoryear{{Majid} et~al.,}{{Majid} et~al.}{2021}]{2021majid}
{Majid} W.~A.,  et~al., 2021, \mn@doi [\apjl] {10.3847/2041-8213/ac1921}, \href {https://ui.adsabs.harvard.edu/abs/2021ApJ...919L...6M} {919, L6}

\bibitem[\protect\citeauthoryear{{Malenta} et~al.,}{{Malenta} et~al.}{2020}]{2020malenta}
{Malenta} M.,  et~al., 2020, in {Pizzo} R.,  {Deul} E.~R.,  {Mol} J.~D.,  {de Plaa} J.,   {Verkouter} H.,  eds,  Astronomical Society of the Pacific Conference Series Vol. 527, Astronomical Society of the Pacific Conference Series. p.~457

\bibitem[\protect\citeauthoryear{{Marsh} et~al.,}{{Marsh} et~al.}{2016}]{2016marsh}
{Marsh} T.~R.,  et~al., 2016, \mn@doi [\nat] {10.1038/nature18620}, \href {https://ui.adsabs.harvard.edu/abs/2016Natur.537..374M} {537, 374}

\bibitem[\protect\citeauthoryear{{McKinnon}}{{McKinnon}}{1996}]{1996mckinnon}
{McKinnon} M.~M.,  1996, in {Johnston} S.,  {Walker} M.~A.,   {Bailes} M.,  eds,  Astronomical Society of the Pacific Conference Series Vol. 105, IAU Colloq. 160: Pulsars: Problems and Progress. p.~253

\bibitem[\protect\citeauthoryear{{McSweeney}, {Smith}, {Bhat}  \& {Wright}}{{McSweeney} et~al.}{2023}]{2023mcsweeney}
{McSweeney} S.~J.,  {Smith} L.,  {Bhat} N.~D.~R.,   {Wright} G.,  2023, \mn@doi [\apj] {10.3847/1538-4357/acdcf2}, \href {https://ui.adsabs.harvard.edu/abs/2023ApJ...952...73M} {952, 73}

\bibitem[\protect\citeauthoryear{{Metzger}, {Berger}  \& {Margalit}}{{Metzger} et~al.}{2017}]{2017metzger}
{Metzger} B.~D.,  {Berger} E.,   {Margalit} B.,  2017, \mn@doi [\apj] {10.3847/1538-4357/aa633d}, \href {https://ui.adsabs.harvard.edu/abs/2017ApJ...841...14M} {841, 14}

\bibitem[\protect\citeauthoryear{{Morello} et~al.,}{{Morello} et~al.}{2019}]{2019morello}
{Morello} V.,  et~al., 2019, \mn@doi [\mnras] {10.1093/mnras/sty3328}, \href {https://ui.adsabs.harvard.edu/abs/2019MNRAS.483.3673M} {483, 3673}

\bibitem[\protect\citeauthoryear{{Morrison} et~al.,}{{Morrison} et~al.}{2023}]{2023morrison}
{Morrison} I.~S.,  et~al., 2023, \mn@doi [\pasa] {10.1017/pasa.2023.15}, \href {https://ui.adsabs.harvard.edu/abs/2023PASA...40...19M} {40, e019}

\bibitem[\protect\citeauthoryear{{Newville}, {Stensitzki}, {Allen}  \& {Ingargiola}}{{Newville} et~al.}{2014}]{2014newville}
{Newville} M.,  {Stensitzki} T.,  {Allen} D.~B.,   {Ingargiola} A.,  2014, {LMFIT: Non-Linear Least-Square Minimization and Curve-Fitting for Python}, \mn@doi{10.5281/zenodo.11813}

\bibitem[\protect\citeauthoryear{{Olausen} \& {Kaspi}}{{Olausen} \& {Kaspi}}{2014}]{2014olausen}
{Olausen} S.~A.,  {Kaspi} V.~M.,  2014, \mn@doi [\apjs] {10.1088/0067-0049/212/1/6}, \href {https://ui.adsabs.harvard.edu/abs/2014ApJS..212....6O} {212, 6}

\bibitem[\protect\citeauthoryear{{Olszanski}, {Mitra}  \& {Rankin}}{{Olszanski} et~al.}{2019}]{2019olszanski}
{Olszanski} T. E.~E.,  {Mitra} D.,   {Rankin} J.~M.,  2019, \mn@doi [\mnras] {10.1093/mnras/stz2172}, \href {https://ui.adsabs.harvard.edu/abs/2019MNRAS.489.1543O} {489, 1543}

\bibitem[\protect\citeauthoryear{{Oppermann}, {Yu}  \& {Pen}}{{Oppermann} et~al.}{2018}]{2018Oppermann}
{Oppermann} N.,  {Yu} H.-R.,   {Pen} U.-L.,  2018, \mn@doi [\mnras] {10.1093/mnras/sty004}, \href {https://ui.adsabs.harvard.edu/abs/2018MNRAS.475.5109O} {475, 5109}

\bibitem[\protect\citeauthoryear{{Ord}, {Tremblay}, {McSweeney}, {Bhat}, {Sobey}, {Mitchell}, {Hancock}  \& {Kirsten}}{{Ord} et~al.}{2019}]{2019ord}
{Ord} S.~M.,  {Tremblay} S.~E.,  {McSweeney} S.~J.,  {Bhat} N.~D.~R.,  {Sobey} C.,  {Mitchell} D.~A.,  {Hancock} P.~J.,   {Kirsten} F.,  2019, \mn@doi [\pasa] {10.1017/pasa.2019.17}, \href {https://ui.adsabs.harvard.edu/abs/2019PASA...36...30O} {36, e030}

\bibitem[\protect\citeauthoryear{{Padmanabh} et~al.,}{{Padmanabh} et~al.}{2023}]{2023Padmanabh}
{Padmanabh} P.~V.,  et~al., 2023, \mn@doi [\mnras] {10.1093/mnras/stad1900}, \href {https://ui.adsabs.harvard.edu/abs/2023MNRAS.524.1291P} {524, 1291}

\bibitem[\protect\citeauthoryear{{Pastor-Marazuela} et~al.,}{{Pastor-Marazuela} et~al.}{2023}]{2023Pastor}
{Pastor-Marazuela} I.,  et~al., 2023, \mn@doi [\aap] {10.1051/0004-6361/202243339}, \href {https://ui.adsabs.harvard.edu/abs/2023A&A...678A.149P} {678, A149}

\bibitem[\protect\citeauthoryear{{Pelisoli} et~al.,}{{Pelisoli} et~al.}{2023}]{2023pelisoli}
{Pelisoli} I.,  et~al., 2023, \mn@doi [Nature Astronomy] {10.1038/s41550-023-01995-x}, \href {https://ui.adsabs.harvard.edu/abs/2023NatAs...7..931P} {7, 931}

\bibitem[\protect\citeauthoryear{{Philippov} \& {Kramer}}{{Philippov} \& {Kramer}}{2022}]{2022philippov}
{Philippov} A.,  {Kramer} M.,  2022, \mn@doi [\araa] {10.1146/annurev-astro-052920-112338}, \href {https://ui.adsabs.harvard.edu/abs/2022ARA&A..60..495P} {60, 495}

\bibitem[\protect\citeauthoryear{{Pilia} et~al.,}{{Pilia} et~al.}{2016}]{2016pilia}
{Pilia} M.,  et~al., 2016, \mn@doi [\aap] {10.1051/0004-6361/201425196}, \href {https://ui.adsabs.harvard.edu/abs/2016A&A...586A..92P} {586, A92}

\bibitem[\protect\citeauthoryear{{Posselt} et~al.,}{{Posselt} et~al.}{2021}]{2021posselt}
{Posselt} B.,  et~al., 2021, \mn@doi [\mnras] {10.1093/mnras/stab2775}, \href {https://ui.adsabs.harvard.edu/abs/2021MNRAS.508.4249P} {508, 4249}

\bibitem[\protect\citeauthoryear{{Price}}{{Price}}{2016}]{2016price}
{Price} D.~C.,  2016, {PyGDSM: Python interface to Global Diffuse Sky Models}, Astrophysics Source Code Library, record ascl:1603.013

\bibitem[\protect\citeauthoryear{{Rajwade} et~al.,}{{Rajwade} et~al.}{2020}]{2021rajwade}
{Rajwade} K.,  et~al., 2020, in {Evans} C.~J.,  {Bryant} J.~J.,   {Motohara} K.,  eds,  Society of Photo-Optical Instrumentation Engineers (SPIE) Conference Series Vol. 11447, Ground-based and Airborne Instrumentation for Astronomy VIII. p. 114470J (\mn@eprint {arXiv} {2103.08410}), \mn@doi{10.1117/12.2559937}

\bibitem[\protect\citeauthoryear{{Rajwade} et~al.,}{{Rajwade} et~al.}{2022}]{2022rajwade}
{Rajwade} K.~M.,  et~al., 2022, \mn@doi [\mnras] {10.1093/mnras/stac446}, \href {https://ui.adsabs.harvard.edu/abs/2022MNRAS.512.1687R} {512, 1687}

\bibitem[\protect\citeauthoryear{{Reddy} et~al.,}{{Reddy} et~al.}{2017}]{2017reddy}
{Reddy} S.~H.,  et~al., 2017, \mn@doi [Journal of Astronomical Instrumentation] {10.1142/S2251171716410117}, \href {https://ui.adsabs.harvard.edu/abs/2017JAI.....641011R} {6, 1641011}

\bibitem[\protect\citeauthoryear{{Rodriguez} et~al.,}{{Rodriguez} et~al.}{2025}]{2025rodriguez}
{Rodriguez} A.~C.,  et~al., 2025, \mn@doi [\pasp] {10.1088/1538-3873/adb0f1}, \href {https://ui.adsabs.harvard.edu/abs/2025PASP..137b4202R} {137, 024202}

\bibitem[\protect\citeauthoryear{{Sanidas}, {Caleb}, {Driessen}, {Morello}, {Rajwade}  \& {Stappers}}{{Sanidas} et~al.}{2018}]{2018sanidas}
{Sanidas} S.,  {Caleb} M.,  {Driessen} L.,  {Morello} V.,  {Rajwade} K.,   {Stappers} B.~W.,  2018, in {Weltevrede} P.,  {Perera} B.~B.~P.,  {Preston} L.~L.,   {Sanidas} S.,  eds, ~ Vol. 337, Pulsar Astrophysics the Next Fifty Years. pp 406--407, \mn@doi{10.1017/S1743921317009310}

\bibitem[\protect\citeauthoryear{{Swarup}, {Ananthakrishnan}, {Kapahi}, {Rao}, {Subrahmanya}  \& {Kulkarni}}{{Swarup} et~al.}{1991}]{1991swarup}
{Swarup} G.,  {Ananthakrishnan} S.,  {Kapahi} V.~K.,  {Rao} A.~P.,  {Subrahmanya} C.~R.,   {Kulkarni} V.~K.,  1991, Current Science, \href {https://ui.adsabs.harvard.edu/abs/1991CSci...60...95S} {60, 95}

\bibitem[\protect\citeauthoryear{{Thorsett}}{{Thorsett}}{1991}]{1991thorsett}
{Thorsett} S.~E.,  1991, \mn@doi [\apj] {10.1086/170355}, \href {https://ui.adsabs.harvard.edu/abs/1991ApJ...377..263T} {377, 263}

\bibitem[\protect\citeauthoryear{{Tingay} et~al.,}{{Tingay} et~al.}{2013}]{2013tingay}
{Tingay} S.~J.,  et~al., 2013, \mn@doi [\pasa] {10.1017/pasa.2012.007}, \href {https://ui.adsabs.harvard.edu/abs/2013PASA...30....7T} {30, e007}

\bibitem[\protect\citeauthoryear{{Tremblay} et~al.,}{{Tremblay} et~al.}{2015}]{2015tremblay}
{Tremblay} S.~E.,  et~al., 2015, \mn@doi [\pasa] {10.1017/pasa.2015.6}, \href {https://ui.adsabs.harvard.edu/abs/2015PASA...32....5T} {32, e005}

\bibitem[\protect\citeauthoryear{{Van Straten} \& {Bailes}}{{Van Straten} \& {Bailes}}{2011}]{2011vanstraten}
{Van Straten} W.,  {Bailes} M.,  2011, \mn@doi [\pasa] {10.1071/AS10021}, \href {https://ui.adsabs.harvard.edu/abs/2011PASA...28....1V} {28, 1}

\bibitem[\protect\citeauthoryear{{Wang}, {Zhang}, {Chen}  \& {Xu}}{{Wang} et~al.}{2019}]{2019wang}
{Wang} W.,  {Zhang} B.,  {Chen} X.,   {Xu} R.,  2019, \mn@doi [\apjl] {10.3847/2041-8213/ab1aab}, \href {https://ui.adsabs.harvard.edu/abs/2019ApJ...876L..15W} {876, L15}

\bibitem[\protect\citeauthoryear{{Wayth} et~al.,}{{Wayth} et~al.}{2018}]{2018wayth}
{Wayth} R.~B.,  et~al., 2018, \mn@doi [\pasa] {10.1017/pasa.2018.37}, \href {https://ui.adsabs.harvard.edu/abs/2018PASA...35...33W} {35, e033}

\bibitem[\protect\citeauthoryear{{Weltevrede}}{{Weltevrede}}{2016}]{2016weltevrede}
{Weltevrede} P.,  2016, \mn@doi [\aap] {10.1051/0004-6361/201527950}, \href {https://ui.adsabs.harvard.edu/abs/2016A&A...590A.109W} {590, A109}

\bibitem[\protect\citeauthoryear{{Yao}, {Manchester}  \& {Wang}}{{Yao} et~al.}{2017}]{ymw16}
{Yao} J.~M.,  {Manchester} R.~N.,   {Wang} N.,  2017, \mn@doi [\apj] {10.3847/1538-4357/835/1/29}, \href {https://ui.adsabs.harvard.edu/abs/2017ApJ...835...29Y} {835, 29}

\bibitem[\protect\citeauthoryear{{Zhao} et~al.,}{{Zhao} et~al.}{2019}]{2019zhao}
{Zhao} R.-S.,  et~al., 2019, \mn@doi [\apj] {10.3847/1538-4357/ab05de}, \href {https://ui.adsabs.harvard.edu/abs/2019ApJ...874...64Z} {874, 64}

\bibitem[\protect\citeauthoryear{{Zheng} et~al.,}{{Zheng} et~al.}{2017}]{2017zheng}
{Zheng} H.,  et~al., 2017, \mn@doi [\mnras] {10.1093/mnras/stw2525}, \href {https://ui.adsabs.harvard.edu/abs/2017MNRAS.464.3486Z} {464, 3486}

\bibitem[\protect\citeauthoryear{{de Ruiter} et~al.,}{{de Ruiter} et~al.}{2024}]{2024deruiter}
{de Ruiter} I.,  et~al., 2024, \mn@doi [arXiv e-prints] {10.48550/arXiv.2408.11536}, \href {https://ui.adsabs.harvard.edu/abs/2024arXiv240811536D} {p. arXiv:2408.11536}

\makeatother
\end{thebibliography}


\bsp	
\label{lastpage}
\end{document}